\documentstyle[12pt,epsf]{article}
\renewcommand{\baselinestretch}{1.2}
\newcommand{\mb}[1]{\mbox{\normalsize\boldmath $#1$}}
\newcommand{\uno}{{ 1\:\!\!\!\mbox{I}}}
\newcommand{\I}{\mb{I}}

\def\npb#1#2#3{    {\it Nucl. Phys. }{\bf B #1} (19#2) #3}
\def\plb#1#2#3{    {\it Phys. Lett. }{\bf B #1} (19#2) #3}
\def\prd#1#2#3{    {\it Phys. Rev. }{\bf D #1} (19#2) #3}
\def\prep#1#2#3{   {\it Phys. Rep. }{\bf #1} (19#2) #3}
\def\prl#1#2#3{    {\it Phys. Rev. Lett. }{\bf #1} (19#2) #3}

\def\ppnp#1#2#3{   {\it Prog. Part. Nucl. Phys. }{\bf #1} (19#2) #3}

\def\zpc#1#2#3{    {\it Zeit. f\"ur Physik }{\bf C #1} (19#2) #3}

\def\diag{\mathop{\mbox{diag}}}

\def\Im{\mathop{\mbox{Im}}}
\def\Re{\mathop{\mbox{Re}}}
\def\Tr{\mathop{\mbox{Tr}}}

\newcommand{\qslash}{/ \mskip-8mu q}
\newcommand{\bea}{\begin{eqnarray}}
\newcommand{\beq}{\begin{equation}}
\newcommand{\eea}{\end{eqnarray}}
\newcommand{\eeq}{\end{equation}}
\newcommand{\nn}{\nonumber}

\newcommand{\als}{\alpha_s}
\newcommand{\db}{\bar{d}}
\newcommand{\gd}{\gamma_\mu}
\newcommand{\gu}{\gamma^\mu}

\begin{document}
\pagestyle{empty} 
\begin{flushright}
ROM2F/96/21 \\
hep-ph/9604387\\
April, 1996 \\
\end{flushright}
\centerline{\Large{\bf{A Complete Analysis of FCNC and CP Constraints}}} 
\centerline{\Large{\bf{in General SUSY Extensions of the Standard Model}}}
\vskip 1cm
\centerline{\bf{ F. Gabbiani$^{a}$, E. Gabrielli$^{b}$,
A. Masiero$^c$ and L. Silvestrini$^{d,b}$}}
\vskip .5cm
\centerline{$^a$ Department of Physics and Astronomy}
\centerline{University of Massachusetts, Amherst, MA 01003, USA}
\centerline{$^b$ INFN, Sezione di Roma II,}
\centerline{Via della Ricerca Scientifica 1, I-00133 Roma, Italy.}
\centerline{$^c$ Dip. di Fisica, Universit\`{a} di Perugia and}
\centerline{INFN, Sezione di Perugia, Via Pascoli, I-06100 Perugia, Italy.}
\centerline{$^d$ Dipartimento di Fisica, Universit\`a di Roma ``Tor Vergata'',}
\centerline{Via della Ricerca Scientifica 1, I-00133 Roma, Italy.}
\begin{abstract}
We analyze the full set of constraints on gluino- and photino-mediated SUSY 
contributions to FCNC and CP violating phenomena. We use the mass insertion 
method, hence providing a model-independent parameterization which can be 
readily applied in testing extensions of the MSSM. In addition to clarifying 
controversial points in the literature, we provide a more exhaustive analysis 
of the CP constraints, in particular concerning 
$\varepsilon^\prime/\varepsilon$.
As physically meaningful applications of our analysis, we study the 
implications in SUSY-GUT's and effective supergravities with flavour 
non-universality. This allows us to detail the domain of applicability and the 
correct procedure of implementation of the FC mass insertion approach.
\end{abstract}
\vskip 3.0cm

\vfill\eject
\pagestyle{empty}\clearpage
\setcounter{page}{1}
\pagestyle{plain}
\newpage
\section{Introduction}  
  In looking for new physics beyond the electroweak Standard Model (SM) it is
useful to regard the SM itself as an effective low energy theory valid up to
some energy scale $\Lambda$ at which the new physics sets in. One is then led
to write all possible operators invariant under  $SU(3)\otimes SU(2) \otimes 
U(1)$ using the
fields of the SM. They can be organized according to their dependence on
$\Lambda$. It is well known that as long as one writes operators not exceeding
dimension four there are crucial conservations which automatically show up:
baryon (B) and lepton (L) numbers and the absence of tree-level
 flavour changing neutral
currents (FCNC). However, as soon as one proceeds beyond dimension four (i.e., 
one considers non-renormalizable operators which are suppressed by powers of
$\Lambda$), these conservations are no longer automatically guaranteed. Either
one has to choose large values for $\Lambda$ (for instance, the grand 
unification or
the Planck scale), or, if $\Lambda$ is assumed to be not so far from the Fermi
scale, additional constraints have to be imposed to play on the safe side in
relation to B, L and FCNC violating processes.

  Low energy supersymmetry (SUSY)~\cite{susygeneral} 
enters this latter class of models with new
 physics close enough to the Fermi scale. The problem of too violent B and L
 violations is more elegantly solved by the imposition of an additional discrete
 symmetry, the R-parity. As for the FCNC issue, given that now we are in the
 presence of new particles, the scalar partners of the fermions (sfermions)
 carrying flavour number, new constraints will have to be imposed to suppress
 operators of dimension greater than four, leading to potentially large FCNC 
rates. They amount to
 very severe limitations on the pattern of the sfermion mass matrices: they must
 be either very close to the unit matrix in flavour space (flavour universality)
 or almost proportional to the corresponding fermion mass matrices (alignment).
 Both universality or alignment can be either preset as kind of ``initial"
 conditions \cite{string, symmetries} or 
result from some dynamics of the theory \cite{plastic, gauge}. 
A very intense work on
 these different options has been going on recently, but, certainly, the flavour
 problem in SUSY remains quite an intriguing issue to be further explored.

  Given a specific SUSY model it is in principle possible to make a full 
 computation of all the FCNC (and, possibly, also CP violating) phenomena in
 that context. This is the case, for instance, of the minimal supersymmetric
 standard model (MSSM) where these detailed computations have led to the result
 of utmost importance that this model succeeds to pass unscathed all the severe
 FCNC and CP tests. However, given the variety of options that exist in
 extending the MSSM (for instance embedding it in some more fundamental theory
 at larger scales), it is important to have a way to extract from the whole
 host
 of FCNC and CP phenomena a set of upper limits on quantities which can be
 readily computed in any chosen SUSY frame. Namely, one needs some kind of 
model-independent parameterization of the FCNC and 
CP quantities in SUSY to have an
accessible immediate test of variants of the MSSM. 

  The best parameterization of this kind that has been proposed is in the
 framework of the so-called mass insertion approximation~\cite{mins}. 
It concerns the most peculiar source of FCNC SUSY contributions that do not 
arise from the mere supersymmetrization of the FCNC in the SM. They originate 
from the FC couplings of gluinos and neutralinos to fermions and 
sfermions~\cite{FCNC}. One chooses a basis
for the fermion and sfermion states where all the couplings of these particles
to neutral gauginos are flavour diagonal, while the FC is exhibited by the
non-diagonality of the sfermion propagators. Denoting by $\Delta$ the
off-diagonal terms in the sfermion mass matrices (i.e. the mass terms relating
sfermion of the same electric charge, but different flavour), the sfermion
propagators can be expanded as a series in terms of $\delta = \Delta/
\tilde{m}^2$
where $\tilde{m}$ is an average sfermion mass to be defined in more detail 
below. 
As long as $\Delta$ is significantly smaller than $\tilde{m}^2$, 
we can just take
the first term of this expansion and, then, the experimental information
concerning FCNC and CP violating phenomena translates into upper bounds on 
these $\delta$'s.
 
  Obviously the above mass insertion method presents the major advantage that
 one does not need the full diagonalization of the sfermion mass matrices to
 perform a test of the SUSY model under consideration in the FCNC sector. It is
 enough to compute ratios of the off-diagonal over the diagonal entries of the
 sfermion mass matrices and compare the results with the general bounds on the
 $\delta$'s that we provide here from all available experimental information.

   There already exist two previous extensive analyses of this kind in the
 literature~\cite{gabbmas, hagelin}. 
Apart from the obvious improvements due to some progress in the
 FCNC and CP experimental results, we were motivated to perform this vast
 analysis by a twofold reason. On one hand, the two previous analyses differed
 in some results and, given the present interest in these kinds of bounds,
it is important to clarify the controversial points.
On the other hand, the quantities related to CP violation
 had not received enough attention in the previous works nor had significant
 flavour-conserving quantities (like the electric dipole moments of the neutron
 and of the electron or the radiative contributions to fermion masses) been
 computed in these analyses.
Moreover, it is of interest to test this
 powerful mass insertion method in ``realistic" sources of potentially large FCNC
 effects in SUSY, namely models with non-universal soft SUSY breaking terms and
 GUT extensions of the MSSM. In the final part of this work we will provide some
 examples of this kind of analyses. Some salient features of the present 
analysis are already contained in a work of ours~\cite{noi}. Here we give a 
more exhaustive treatment of the previous results and we enlarge our study to 
include physically relevant applications.

It is important to keep in mind that our analysis focuses only on the gluino- 
or photino-mediated FCNC contributions. It is well known that other SUSY 
sectors can yield relevant (and, sometimes, even dominant) contributions to 
FCNC. We refer to chargino or charged Higgs exchanges, in particular. However, 
the full computation of all these SUSY effects requires the full specification 
of the model. On the contrary, as we said, the spirit of our study is to 
provide a model-independent way to make a first check of the FCNC impact on 
classes of SUSY theories.

  Before closing this introductory part, we provide some more details on the
 $\delta$ quantities on which we plan to set limits and on the procedure 
that is actually
 followed to derive such bounds.

  There exist four different 
$\Delta$ mass insertions connecting flavours $i$ and $j$
 along a sfermion propagator: $\left(\Delta_{ij}\right)_{LL}$, 
$\left(\Delta_{ij}\right)_{RR}$, $\left(\Delta_{ij}\right)_{LR}$ and 
$\left(\Delta_{ij}\right)_{RL}$. The indices $L$ and $R$ refer to the 
helicity of 
the
 fermion partners. The size of these $\Delta$'s can be quite different. For
 instance, it is well known that in the MSSM case, only the $LL$ mass insertion
 can change flavour, while all the other three above mass insertions are flavour
 conserving, i.e. they have $i=j$. In this case to realize a $LR$ or $RL$ 
flavour
 change one needs a double mass 
insertion with the flavour changed solely in a $LL$
 mass insertion and a subsequent flavour-conserving $LR$ mass insertion. 
Even worse
 is the case of a FC $RR$ transition: in the MSSM this can be accomplished only
 through a laborious set of three mass insertions, two flavour-conserving $LR$
transitions and an $LL$ FC insertion. Notice also that generally the 
$\Delta_{LR}$
quantity does not necessarily coincide with $\Delta_{RL}$. For instance, in the
MSSM and in several other cases, one flavour-conserving mass insertion is
proportional to the mass of the corresponding right-handed fermion. Hence ,
$(\Delta_{ij})_{LR}$ and $(\Delta_{ij})_{RL}$ are proportional to the mass of 
the $i$-th
and $j$-th fermion, respectively. We will comment further on this point later on
in the paper.  Instead of the dimensional quantities $\Delta$ it is more
useful to provide bounds making use of dimensionless quantities, $\delta$, 
that are obtained dividing the mass insertions by an average sfermion mass.

Concerning the definition of the average sfermion mass, in most cases it is
indifferent how one exactly defines it (for instance, one can take half the sum
of the two diagonal entries for the $i$-th and $j$-th sfermion). However, 
care must
be taken when the degree of degeneracy of the sfermion is not so high (obviously
in this case one has to check carefully that the whole mass insertion approach
makes sense). Making quantitative checks it turned out that we were reproducing
more closely the exact result, that is obtained by fully diagonalizing the 
sfermion
mass matrices, by taking as a definition of average mass between the sfermions
$m_1$ and $m_2$ the quantity $\sqrt{m_1m_2}$.

  Finally an important point to underline is what we actually mean by using the
 experimental data in the derivation of the upper bounds on the $\delta$'s.
 Clearly we have to take into account also the theoretical uncertainty in the SM
 in evaluating the different processes. For instance when considering the
 process $b \to s \gamma$ to 
derive the bounds on the $\delta_{23}$ quantities we
 consider as ``experimental" input the interval $1\times 10^{-4} < \mbox{BR}
(b\to s \gamma)<4\times 10^{-4}$,
which takes into account the still wide theoretical uncertainty in evaluating
this BR within the SM. In any case, to avoid any ambiguity, for each process
under consideration we will explicitly provide the ``experimental" input 
that we use.

  The paper is organized as follows. In Sect.~\ref{df2} we analyze the bounds 
on the
  $\delta$'s coming from $\Delta F=2$ processes. After briefly summarizing the
  method of the effective hamiltonian that we make use of, we construct such
  effective hamiltonian for $\Delta S=2$ transitions and we derive the 
corresponding bounds
  on the $\delta_{12}$ quantities for both the CP conserving and CP violating
  cases. We extend this analysis also to the D and B systems. 
Section~\ref{df1} follows
  the same pattern of analysis as the previous one for the $\Delta F=1$
  processes. In particular we devote special attention to the evaluation of the
  bounds coming from the CP violating $\varepsilon^\prime$ quantity. 
A comment on the
  role played by the QCD corrections in our analysis is provided in 
Sect.~\ref{sec:qcd}.
Then Sect.~\ref{sec:models} compares the bounds that we derived for some 
$\delta$ 
quantities
with their values in two classes of extensions of the MSSM which are actively
investigated at the moment and that present important sources of FCNC. We refer
to SUSY GUT's where not only are such large contributions to FCNC potentially
present, but, at least if we can reliably trust the use of the Renormalization 
Group Equations (RGE's) in the
evolution from the Planck to the GUT scale, they are indeed unavoidably there
with important constraints on any such extension. The second class is given by
models where the initial SUSY soft breaking terms lack the property of flavour
universality. Although one can propose specific situations of SUSY breaking
where these embarrassingly high contributions to FCNC are absent, it is certainly
true that in the general case of supergravity models derived from four
dimensional strings one expects such non-universality to be indeed present.
Finally some conclusions and an outlook are presented in Sect.~\ref{concl}.

\section{$\Delta F=2$ processes}
\label{df2}
In this Section we will describe the calculation of the amplitude for 
$\Delta F=2$ processes and we will present the phenomenological 
analysis of these transitions and the limits on sfermion masses thereby 
obtained.
\subsection{Effective Hamiltonian}
\label{EH}
Let us briefly recall the procedure to calculate the Effective 
Hamiltonian (EH) for a given process. One has to go through the following 
steps:
\begin{enumerate}
	\item  calculate the amplitude between quark and gluon 
	states of definite momenta in the full theory;
	
	\item  choose a basis of local operators for the effective theory and 
	calculate their matrix elements between the same states used in point 1;
	
	\item  determine the coefficients of the operators in the EH by matching 
	the full theory with the effective one.
\end{enumerate}
The matching is given by the following relation:
\begin{equation}
	\langle f \vert S \vert i \rangle = -i\sum_{j} C_j \langle f \vert O_j 
	\vert i \rangle \; ,
	\label{match}
\end{equation}
where $C_i$ are the Wilson coefficients and $O_i$ the operators of the EH:
\begin{equation}
\cal{H}_{eff}=\sum_{i}C_{i}O_{i}
\end{equation}

\begin{figure}     
\epsfysize=15cm 
\epsfxsize=17cm 
\epsffile{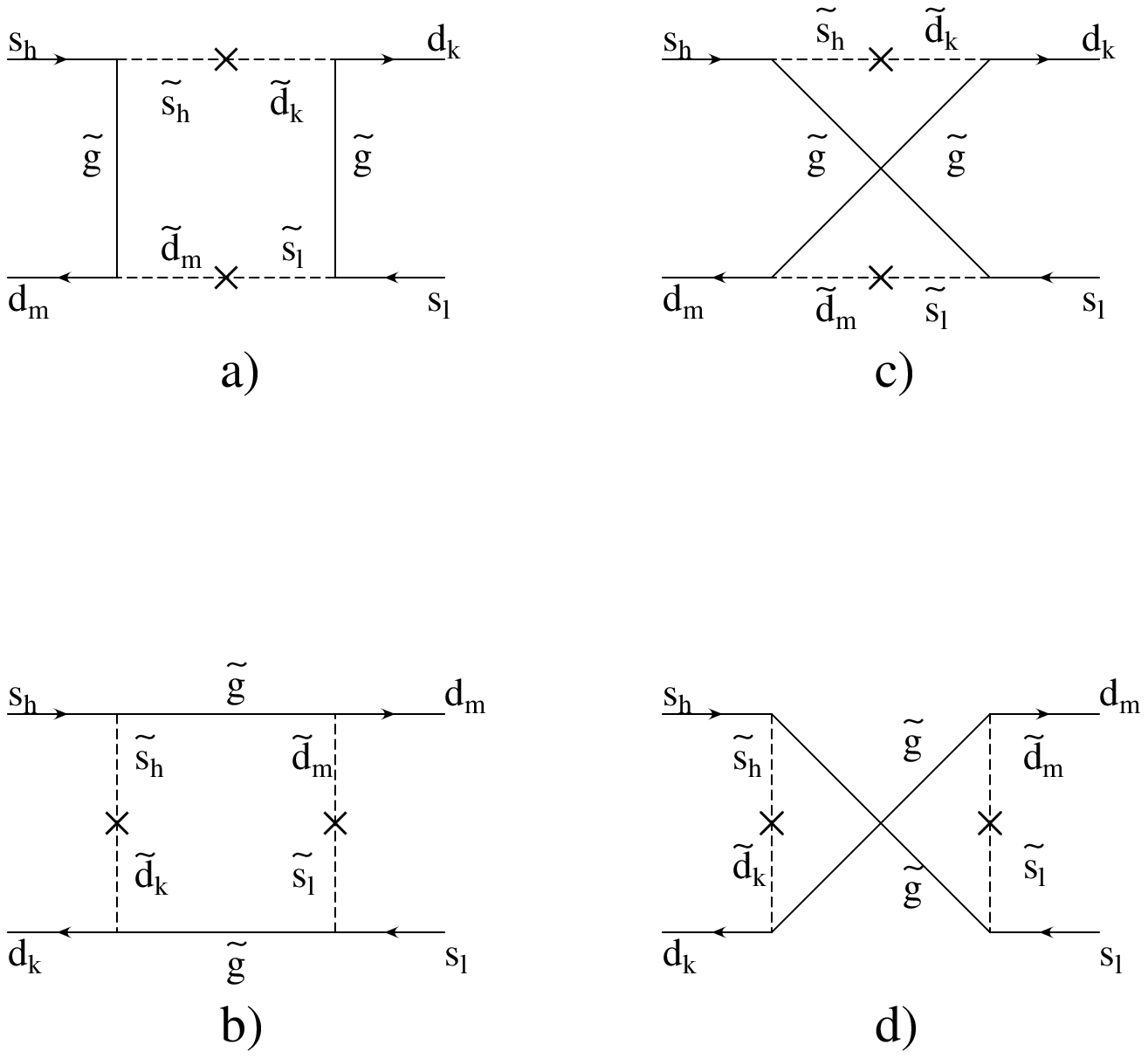} 
\caption{Feynman diagrams for $\Delta S=2$ transitions, with $h,k,l,m=\{L,R\}$.}
\label{figds2}
\end{figure}

Let us now specialize to the case of $\Delta S=2$ processes. The 
amplitude in the full theory is given by the calculation of the diagrams 
in fig.~\ref{figds2}. 
Having calculated the amplitude in the full theory, we now have to 
choose a basis of local operators and perform the matching. 
Our choice is the following:
\begin{eqnarray}
	Q_1 & = & \db^{\alpha}_L \gd s^{\alpha}_L \db^{\beta}_{L} \gu s^{\beta}_L\; ,
	\nonumber \\
	Q_2 & = & \db^{\alpha}_R  s^{\alpha}_L \db^{\beta}_R s^{\beta}_L\; ,
	\nonumber \\
	Q_3 & = & \db^{\alpha}_R  s^{\beta}_L \db^{\beta}_R s^{\alpha}_L\; ,
	\nonumber \\
	Q_4 & = & \db^{\alpha}_R  s^{\alpha}_L \db^{\beta}_L s^{\beta}_R\; ,
	\nonumber \\
	Q_5 & = & \db^{\alpha}_R  s^{\beta}_L \db^{\beta}_L s^{\alpha}_R\; ,
	\label{Qi}
\end{eqnarray}
plus the operators $\tilde{Q}_{1,2,3}$ obtained from the $Q_{1,2,3}$ by the 
exchange $ L \leftrightarrow R$. 
Here $q_{R,L}=\frac{(1\pm\gamma_5)}{2}q$, and $\alpha$ and $\beta$ are colour
indices.
The colour matrices normalization is $\Tr(t^A t^B)= \delta^{AB}/2$.

Performing the matching we obtain 
the following result for the $\Delta S=2$ EH:
\begin{eqnarray}
	\cal{H}_{eff}&=&-\frac{\als^2}{216 m_{\tilde{q}}^2}\Biggl\{  
	\left(\delta^d_{12}\right)^2_{LL}  
	\left(  24\,Q_1\,x\,f_6(x) + 66\,Q_1\,\tilde{f}_6(x) \right)
	\nonumber \\
	&+&  \left(\delta^d_{12}\right)^2_{RR}   
	\left(  24\,\tilde{Q}_1\,x\,f_6(x) + 66\,\tilde{Q}_1\,
	\tilde{f}_6(x) \right)
	\nonumber \\
	&+&  \left(\delta^d_{12}\right)_{LL}\left(\delta^d_{12}\right)_{RR}  
	\left( 504\,Q_4\,x\,f_6(x) - 72\,Q_4\,\tilde{f}_6(x) 
	\right. \nonumber \\
	&&+ \left. 24\,Q_5\,x\,f_6(x) + 120\,Q_5\,\tilde{f}_6(x) \right)
	\nonumber \\
	&+&  \left(\delta^d_{12}\right)^2_{RL}  
	\left(  204\,Q_2\,x\,f_6(x) - 36\,Q_3\,x\,f_6(x) \right)
	\nonumber \\
	&+&  \left(\delta^d_{12}\right)^2_{LR}  
	\left(  204\,\tilde{Q}_2\,x\,f_6(x) - 36\,\tilde{Q}_3\,x\,f_6(x) \right)
	\nonumber \\
	&+&  \left(\delta^d_{12}\right)_{LR}\left(\delta^d_{12}\right)_{RL}  
	\left( - 132\,Q_4\,\tilde{f}_6(x) - 180\,Q_5\,\tilde{f}_6(x)
	\right)  \Biggr\},
	\label{heds2}
\end{eqnarray}
where $x=m^2_{\tilde{g}}/m_{\tilde{q}}^2$, $m_{\tilde{q}}$ 
is the average squark mass, 
$m_{\tilde{g}}$ is the gluino mass and the functions $f_6(x)$ and 
$\tilde{f}_6(x)$ are given by (we follow the notation of ref.~\cite{hagelin}):
\begin{eqnarray}
f_6(x)=\frac{6(1+3x)\ln x +x^3-9x^2-9x+17}{6(x-1)^5}\; , \nonumber  \\
\tilde{f}_6(x)=\frac{6x(1+x)\ln x -x^3-9x^2+9x+1}{3(x-1)^5}\; . 
\end{eqnarray}

This result is in agreement with ref.~\cite{gerard}, but differs from 
ref.~\cite{hagelin} in the Left-Right terms.
In order to clarify these discrepancies, we explicitly give the contribution 
to the EH coming from the  
diagrams in fig.~\ref{figds2} in the case of Left-Right mass 
insertions:
\begin{eqnarray}
a)+b)&=&\frac{\als^2}{216 m_{\tilde{q}}^2} \Biggl\{ \left[ 
\left(\delta^d_{12}\right)^2_{RL}
\left( - 252\,Q_2\,x\,f_6(x) - 12\,Q_3\,x\,f_6(x) \right) +
L \leftrightarrow R  
\right] \nonumber \\
&+& 
\left(\delta^d_{12}\right)_{LR}
\left(\delta^d_{12}\right)_{RL}
\left( 12\,Q_4\,\tilde{f}_6(x) + 252\,Q_5\,\tilde{f}_6(x) \right) 
\Biggr\} ,
\nonumber \\
c)+d)&=&\frac{\als^2}{216 m_{\tilde{q}}^2}
\Biggl\{ \left[ \left(\delta^d_{12}\right)^2_{RL} 
\left(48\,Q_2\,x\,f_6(x) + 48\,Q_3\,x\,f_6(x)  \right) +
L \leftrightarrow R
\right] 
\nonumber \\
&+& 
\left(\delta^d_{12}\right)_{LR} 
\left(\delta^d_{12}\right)_{RL} 
\left(  120\,Q_4\,\tilde{f}_6(x) - 72\,Q_5\,\tilde{f}_6(x) \right) 
\Biggr\}.
\end{eqnarray}   
Our results also differ from ref.~\cite{gabbmas}. 
To obtain our results,
in eqs. (3.2 a) and c) of ref. \cite{gabbmas} the terms 
proportional to
the function $M(x)$ must be multiplied by the coefficient ($-1/2$), 
while in eq. 
(3.2 b) the function $G(x)$ must be multiplied by ($-1$).

\subsection{Hadronic Matrix Elements}
We now give the matrix elements of the operators $Q_i$ between $K$ 
mesons, in the vacuum insertion approximation (VIA).
From PCAC we obtain the basic formulas:
\begin{eqnarray}
	&\,& \langle K^{0} \vert \db^{\alpha} \gd \gamma_{5} s^{\alpha} 
	\vert 0 \rangle \langle 0 
	\vert \db^{\beta} \gu \gamma_{5} s^{\beta} 
	\vert \bar{K}^{0} \rangle  = 
	\frac{m_{K}f_{K}^{2}}{2}
	\nonumber \\
	&\,& \langle K^{0} \vert \db^{\alpha} \gamma_{5} s^{\alpha} \vert 0 
	\rangle \langle 0 
	\vert \db^{\beta} \gamma_{5} s^{\beta}
	\vert \bar{K}^{0} \rangle = -\frac{m_{K}f_{K}^{2}}{2} \left(
	\frac{m_{K}}{m_{s}+m_{d}}\right)^{2} \; ,
	\label{me1}
\end{eqnarray}
where $m_{K}$ is the mass of the $K$ mesons and $m_{s}$, $m_{d}$ are the 
masses of $s$ and $d$ quarks respectively. From eq.~(\ref{me1}) we derive:
\begin{eqnarray}
	\langle K^{0} \vert Q_{1} 
	\vert \bar{K}^{0} \rangle & = & \frac{1}{3}m_{K}f_{K}^{2}\; ,
	\nonumber \\
	\langle K^{0} \vert Q_{2} 
	\vert \bar{K}^{0} \rangle & = & -\frac{5}{24}  
	\left(\frac{m_{K}}{m_{s}+m_{d}}\right)^{2}m_{K}f_{K}^{2}\; ,
	\nonumber \\
	\langle K^{0} \vert Q_{3} 
	\vert \bar{K}^{0} \rangle & = & \frac{1}{24} 
	\left(\frac{m_{K}}{m_{s}+m_{d}}\right)^{2}m_{K}f_{K}^{2}\; ,
	\nonumber \\
	\langle K^{0} \vert Q_{4} 
	\vert \bar{K}^{0} \rangle & = & \left[\frac{1}{24} +
	\frac{1}{4} 
	\left(\frac{m_{K}}{m_{s}+m_{d}}\right)^{2}\right]m_{K}f_{K}^{2}\; ,
	\nonumber \\
	\langle K^{0} \vert Q_{5}
	\vert \bar{K}^{0} \rangle & = &\left[\frac{1}{8} +
	\frac{1}{12} 
	\left(\frac{m_{K}}{m_{s}+m_{d}}\right)^{2}\right] m_{K}f_{K}^{2}\; ,
	\label{meds2}
\end{eqnarray}
where we have set equal to one all the $B$ parameters.  
Our results for the matrix elements in eq.~(\ref{meds2}) agree with 
ref.~\cite{gerard} but disagree with ref.~\cite{hagelin} in the sign of the 
terms proportional to $(m_{K}/(m_{s} + m_{d}))^{2}$. 
This is probably due to a difference in the sign of the 
last line of eq.~(\ref{me1}). We have checked that the sign we have 
obtained agrees with other previous calculations (see e.g. ref.~\cite{soni}). 

\subsection{General Analysis of $\Delta F =2$ Processes}
We now present the results of a model-independent analysis of low energy
$\Delta F =2$ processes. Let us start from $K^{0} - \bar{K}^{0}$ mixing.
The $K_{L}-K_{S}$ mass difference $\Delta m_{K}$ is given by:
\begin{equation}
	\Delta m_{K} = 2 \Re \langle K^{0} \vert \cal{H}_{eff} 
	\vert \bar{K}^{0} \rangle \; .
	\label{defdmk}
\end{equation}
Substituting the expressions (\ref{heds2}) for the EH and 
(\ref{meds2}) for the matrix 
elements into (\ref{defdmk}), we obtain the following expression for 
$\Delta m_{K}$:
\begin{eqnarray}
	\Delta m_{K}&=&-\frac{\als^2}{216 m_{\tilde{q}}^2}\frac{2}{3}m_{K}f_{K}^{2}
	\Biggl\{  \left(\delta^d_{12}\right)^2_{LL}  
	\left(  24\,x\,f_6(x) + 66\,\tilde{f}_6(x) \right)
	\nonumber \\
	&+&  \left(\delta^d_{12}\right)^2_{RR}   
	\left(  24\,x\,f_6(x) + 66\,\tilde{f}_6(x) \right)
	\nonumber \\
	&+&  \left(\delta^d_{12}\right)_{LL}\left(\delta^d_{12}\right)_{RR}  
	\left[ \left(384\left(\frac{m_{K}}{m_{s}+m_{d}}\right)^{2} + 72 \right)
	\,x\,f_6(x)  \right. \nonumber \\
	&&+ \left. \left(-24\left(\frac{m_{K}}{m_{s}+m_{d}}\right)^{2} 
	+ 36\right) \tilde{f}_6(x) \right]
	\nonumber \\
	&+&  \left(\delta^d_{12}\right)^2_{LR}  
	\left[  -132 \left(\frac{m_{K}}{m_{s}+m_{d}}\right)^{2} x\,f_6(x) \right]
	\nonumber \\
	&+&  \left(\delta^d_{12}\right)^2_{RL}  
	\left[  -132 \left(\frac{m_{K}}{m_{s}+m_{d}}\right)^{2} x\,f_6(x) \right]
	\nonumber \\
	&+&  \left(\delta^d_{12}\right)_{LR}\left(\delta^d_{12}\right)_{RL}  
	\left[ -144 \left(\frac{m_{K}}{m_{s}+m_{d}}\right)^{2} -84 \right] 
	\tilde{f}_6(x)
	\Biggr\}.
	\label{dmk}
\end{eqnarray}

\begin{figure}   
    \begin{center}
    \epsfysize=8truecm 
    \leavevmode\epsffile{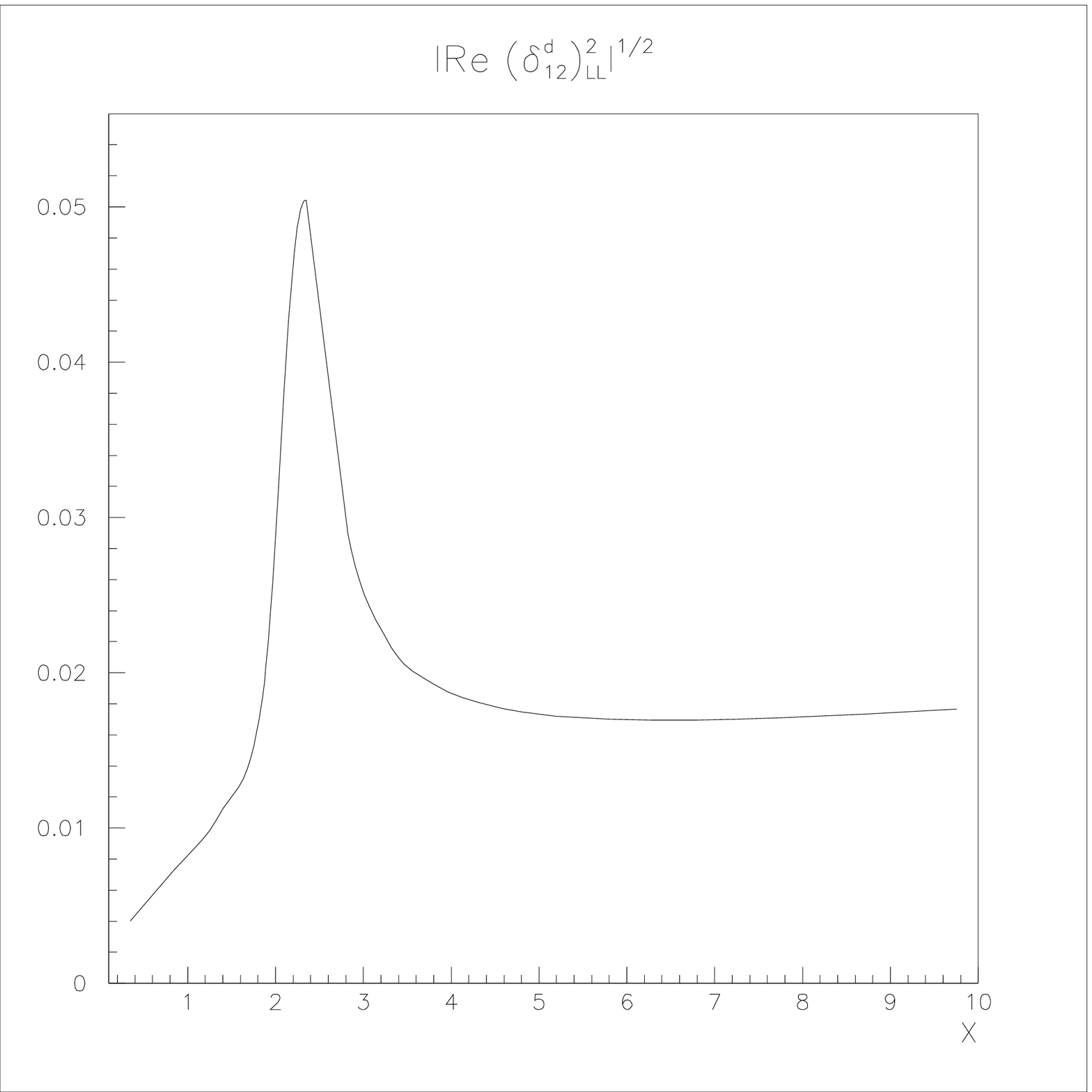}
    \end{center}
    \caption[]{The $\sqrt{\left|\Re  \left(\delta^{d}_{12} 
     \right)_{LL}^{2}\right|} $ as a function
     of $x=m_{\tilde{g}}^2/m_{\tilde{q}}^2$, for  an average squark mass 
     $m_{\tilde{q}}=100\mbox{GeV}$.}
\label{kkll}
\end{figure}

\begin{figure}   
    \begin{center}
    \epsfysize=8truecm 
    \leavevmode\epsffile{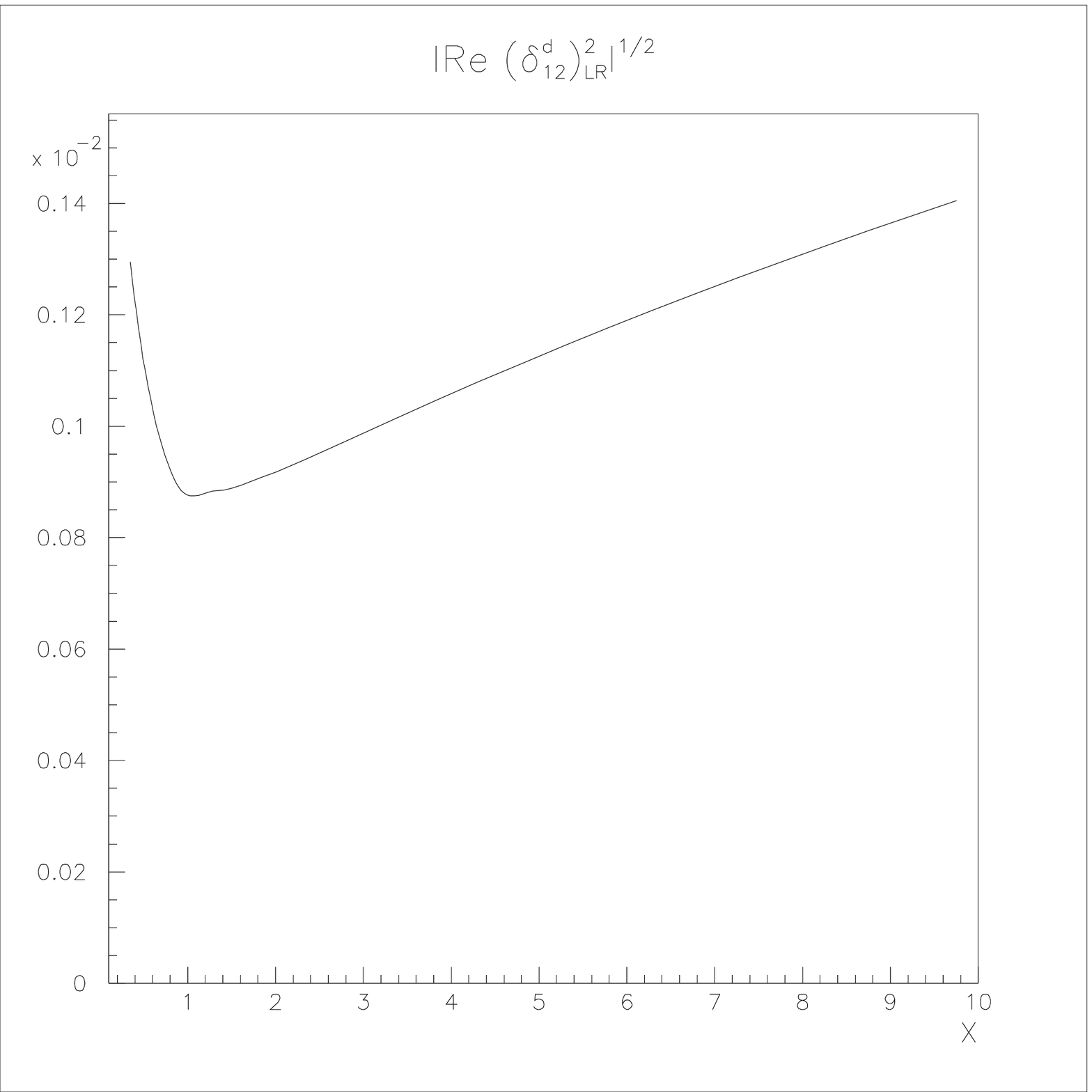}
    \end{center}
    \caption[]{The $\sqrt{\left|\Re  \left(\delta^{d}_{12} 
\right)_{LR}^{2}\right|} $ as a function
     of $x=m_{\tilde{g}}^2/m_{\tilde{q}}^2$, for  an average squark mass 
     $m_{\tilde{q}}=100\mbox{GeV}$.}
\label{kklr}
\end{figure}

\begin{figure}   
    \begin{center}
    \epsfysize=8truecm 
    \leavevmode\epsffile{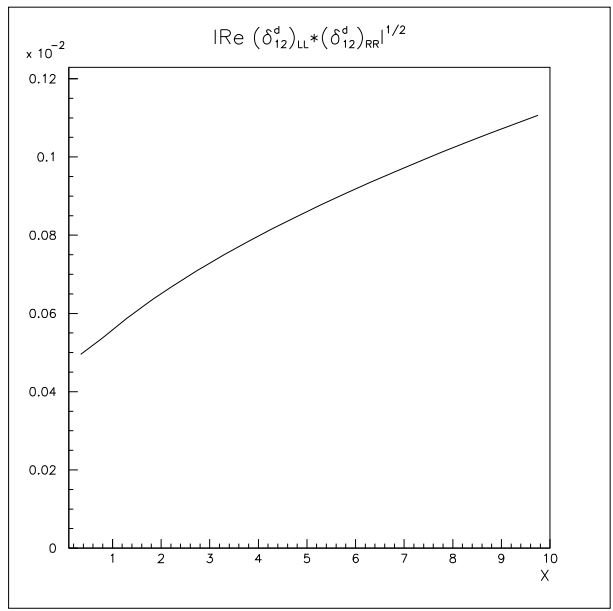}
    \end{center}
    \caption[]{The $\sqrt{\left|\Re  \left(\delta^{d}_{12} 
\right)_{LL} \left(\delta^{d}_{12} 
\right)_{RR}\right|} $ as a function
     of $x=m_{\tilde{g}}^2/m_{\tilde{q}}^2$, for  an average squark mass 
     $m_{\tilde{q}}=100\mbox{GeV}$.}
\label{kkllrr}
\end{figure}

Analogous expressions can be found for $D - \bar{D}$ and $B - \bar{B}$ 
mixing. Starting from eq.~(\ref{dmk}), and imposing that the contribution of 
each term in (\ref{dmk}) does not exceed (in absolute value) the measured 
$\Delta m_{K}$, we obtain the limits on the $\delta$'s reported in 
table~\ref{reds2}, barring accidental cancellations.
Here and in the following we take 
$(\delta_{ij})_{LR}\simeq (\delta_{ij})_{RL}$, to simplify the 
analysis\footnote{As we said in the Introduction, this approximation 
does not hold true in general: 
for example, in the MSSM we 
have $(\delta^d_{12})_{RL}/(\delta^d_{12})_{LR}=m_{d}/m_{s}\ll1$. 
However, as the amplitudes we study are Left-Right symmetric, the bounds that 
we find can be easily extended to the asymmetric case 
$(\delta_{ij})_{LR}\gg (\delta_{ij})_{RL}$ or $(\delta_{ij})_{LR}\ll 
(\delta_{ij})_{RL}$.}.
The  parameters and upper limits used here and in the following 
are reported in tables~\ref{parameters} and \ref{ul}.
In the $K-\bar{K}$ system, limits can also be obtained  from the 
CP-violating parameter $\varepsilon$: these are reported in table~\ref{imds2}.
The dependence of the limits in table~\ref{reds2}  on $x$ is 
given in figures~\ref{kkll}, \ref{kklr} and \ref{kkllrr}. The dependence of 
the limits in table~\ref{imds2} on $x$ is identical.
\begin{table}
 \begin{center}
 \begin{tabular}{||c|c|c|c||}  \hline \hline
 $x$ & $\sqrt{\left|\Re  \left(\delta^{d}_{12} \right)_{LL}^{2}\right|} $ 
 &
 $\sqrt{\left|\Re  \left(\delta^{d}_{12} \right)_{LR}^{2}\right|} $ &
 $\sqrt{\left|\Re  \left(\delta^{d}_{12} \right)_{LL}\left(\delta^{d}_{12}
 \right)_{RR}\right|} $ \\
 \hline
 $
   0.3
 $ &
 $
1.9\times 10^{-2}
 $ & $
7.9\times 10^{-3}
 $ & $
2.5\times 10^{-3}
 $ \\
 $
   1.0
 $ &
 $
4.0\times 10^{-2}
 $ & $
4.4\times 10^{-3}
 $ & $
2.8\times 10^{-3}
 $ \\
 $
   4.0
 $ &
 $
9.3\times 10^{-2}
 $ & $
5.3\times 10^{-3}
 $ & $
4.0\times 10^{-3}
 $ \\ \hline \hline
 $x$ & $\sqrt{\left|\Re  \left(\delta^{d}_{13} \right)_{LL}^{2}\right|} $ 
 &
 $\sqrt{\left|\Re  \left(\delta^{d}_{13} \right)_{LR}^{2}\right|} $ &
 $\sqrt{\left|\Re  \left(\delta^{d}_{13} \right)_{LL}\left(\delta^{d}_{13}
 \right)_{RR}\right|} $ \\
 \hline
 $
   0.3
 $ &
 $
4.6\times 10^{-2}
 $ & $
5.6\times 10^{-2}
 $ & $
1.6\times 10^{-2}
 $ \\
 $
   1.0
 $ &
 $ 
9.8\times 10^{-2}
 $ & $
3.3\times 10^{-2}
 $ & $
1.8\times 10^{-2}
 $ \\
 $
   4.0
 $ &
 $
2.3\times 10^{-1}
 $ & $
3.6\times 10^{-2}
 $ & $
2.5\times 10^{-2}
 $ \\ \hline \hline
 $x$ & $\sqrt{\left|\Re  \left(\delta^{u}_{12} \right)_{LL}^{2}\right|} $ 
 &
 $\sqrt{\left|\Re  \left(\delta^{u}_{12} \right)_{LR}^{2}\right|} $ &
 $\sqrt{\left|\Re  \left(\delta^{u}_{12} \right)_{LL}\left(\delta^{u}_{12}
 \right)_{RR}\right|} $ \\
 \hline
 $
   0.3
 $ &
 $
4.7\times 10^{-2}
 $ & $
6.3\times 10^{-2}
 $ & $
1.6\times 10^{-2}
 $ \\
 $
   1.0
 $ &
 $
1.0\times 10^{-1}
 $ & $
3.1\times 10^{-2}
 $ & $
1.7\times 10^{-2}
 $ \\
 $
   4.0
 $ &
 $
2.4\times 10^{-1}
 $ & $
3.5\times 10^{-2}
 $ & $
2.5\times 10^{-2}
 $ \\ \hline \hline
 \end{tabular}
 \caption[]{Limits on $\mbox{Re}\left(\delta_{ij}\right)_{AB}\left(
\delta_{ij}\right)_{CD}$, with $A,B,C,D=(L,R)$, for an average squark mass
 $m_{\tilde{q}}=500\mbox{GeV}$ and for different values of 
 $x=m_{\tilde{g}}^2/m_{\tilde{q}}^2$. For different values of $m_{\tilde{q}}$, 
 the limits can be obtained multiplying the ones in the table by 
 $m_{\tilde{q}}(\mbox{GeV})/500$.}
 \label{reds2}
 \end{center}
 \end{table}
 \begin{table}
 \begin{center}
 \begin{tabular}{||c|c||}
	\hline \hline
	Constants & Values  \\
	\hline
	$m_{\pi}$ & $140$ MeV  \\
	\hline
	$m_{K}$ & $490$ MeV  \\
	\hline
	$m_{B}$ & $5.278$ GeV  \\
	\hline
	$m_{D}$ & $1.86$ GeV  \\
	\hline
	$f_{\pi}$ & $132$ MeV  \\
	\hline
	$f_{K}$ & $160$ MeV  \\
	\hline
	$f_{B}$ & $200$ MeV \\
	\hline
	$f_{D}$ & $200$ MeV  \\
	\hline
	Re$A_{0}$ & $2.7 \times 10^{-7}$ GeV  \\
	\hline
	$\omega$ & $0.045$  \\
	\hline
	$\tau_{B}$ & $1.49\times 10^{-12}$ s \\
	\hline
	$m_{s}$ & $150$ MeV  \\
	\hline
	$m_{c}$ & $1.5$ GeV  \\
	\hline
	$m_{b}$ & $4.5$ GeV  \\
	\hline
	$\als(M_{W})$ & $0.12$ \\
	\hline
	$\vert \mb{V}_{23} \vert$ & $0.04$ \\
	\hline
	$\vert \mb{V}_{31} \vert$ & $0.01$ \\
	\hline \hline
 \end{tabular}
 \caption[]{Constants used in the phenomenological analysis.}
 \label{parameters}
 \end{center}
 \end{table}
 \begin{table}
 \begin{center}
 \begin{tabular}{||c|c||}
	\hline \hline
	Quantity & Value  \\
	\hline
	$\Delta m_{K}$ & $< 3.521 \times 10^{-12}$ MeV  \\
	\hline
	$\varepsilon$ & $< 2.268 \times 10^{-3}$   \\
	\hline
	$\varepsilon^\prime/\varepsilon$ & $< 2.7 \times 10^{-3}$   \\
	\hline
	$\Delta m_{B}$ & $< 3.75 \times 10^{-10}$ MeV \\
	\hline 
	$\Delta m_{D}$ & $< 1.32 \times 10^{-10}$ MeV \\
	\hline 
	$\mbox{BR}(b \to s \gamma)$ & $(1-4) \times 10^{-4}$ \\
	\hline	
	$\mbox{BR}(\mu \to e \gamma)$ & $<4.9 \times 10^{-11}$ \\
	\hline	
	$\mbox{BR}(\tau \to e \gamma)$ & $<1.2 \times 10^{-4}$ \\
	\hline	
	$\mbox{BR}(\tau \to \mu \gamma)$ & $<4.2 \times 10^{-6}$ \\
	\hline	
	$d_n$ & $<11 \times 10^{-26}\; \;e\;$cm \\
	\hline	
	$d_e$ & $<7 \times 10^{-27}\; \;e\;$cm \\
	\hline \hline
 \end{tabular}
 \caption[]{Limits used in the phenomenological analysis.}
 \label{ul}
 \end{center}
 \end{table}
\begin{table}
 \begin{center}
 \begin{tabular}{||c|c|c|c||}  \hline \hline
  $x$ &
 ${\scriptstyle\sqrt{\left|\Im  \left(\delta^{d}_{12} \right)_{LL}^{2}
\right|} }$ &
 ${\scriptstyle\sqrt{\left|\Im  \left(\delta^{d}_{12} \right)_{LR}^{2}
\right|} }$ &
 ${\scriptstyle\sqrt{\left|\Im  \left(\delta^{d}_{12} \right)_{LL}\left(\delta^{d}_{12}
 \right)_{RR}\right|} }$ \\
 \hline
 $
   0.3
 $ &
 $
1.5\times 10^{-3}
 $ & $
6.3\times 10^{-4}
 $ & $
2.0\times 10^{-4}
 $ \\
 $
   1.0
 $ &
 $
3.2\times 10^{-3}
 $ & $
3.5\times 10^{-4}
 $ & $
2.2\times 10^{-4}
 $ \\
 $
   4.0
 $ &
 $
7.5\times 10^{-3}
 $ & $
4.2\times 10^{-4}
 $ & $
3.2\times 10^{-4}
 $ \\ \hline \hline
 \end{tabular}
 \ \caption[]{Limits on 
$\mbox{Im}\left(\delta_{12}^{d}\right)_{AB}\left(\delta_{12}^{d}\right)_{CD}$, 
with $A,B,C,D=(L,R)$, for 
 an average squark mass $m_{\tilde{q}}=500\mbox{GeV}$ and for different values of 
 $x=m_{\tilde{g}}^2/m_{\tilde{q}}^2$. For different values of $m_{\tilde{q}}$, 
 the limits can be obtained multiplying the ones in the table by 
 $m_{\tilde{q}}(\mbox{GeV})/500$.}
 \label{imds2}
 \end{center}
 \end{table}
\section{$\Delta F=1$ Processes}
\label{df1}
In this Section we will discuss the calculation of the EH for $\Delta 
F=1$ transitions and the phenomenological analysis of these processes, 
following the outline of Section~\ref{df2}.
\subsection{Effective Hamiltonian}
\begin{figure}     
\epsfysize=19cm 
\epsfxsize=18cm 
\epsffile{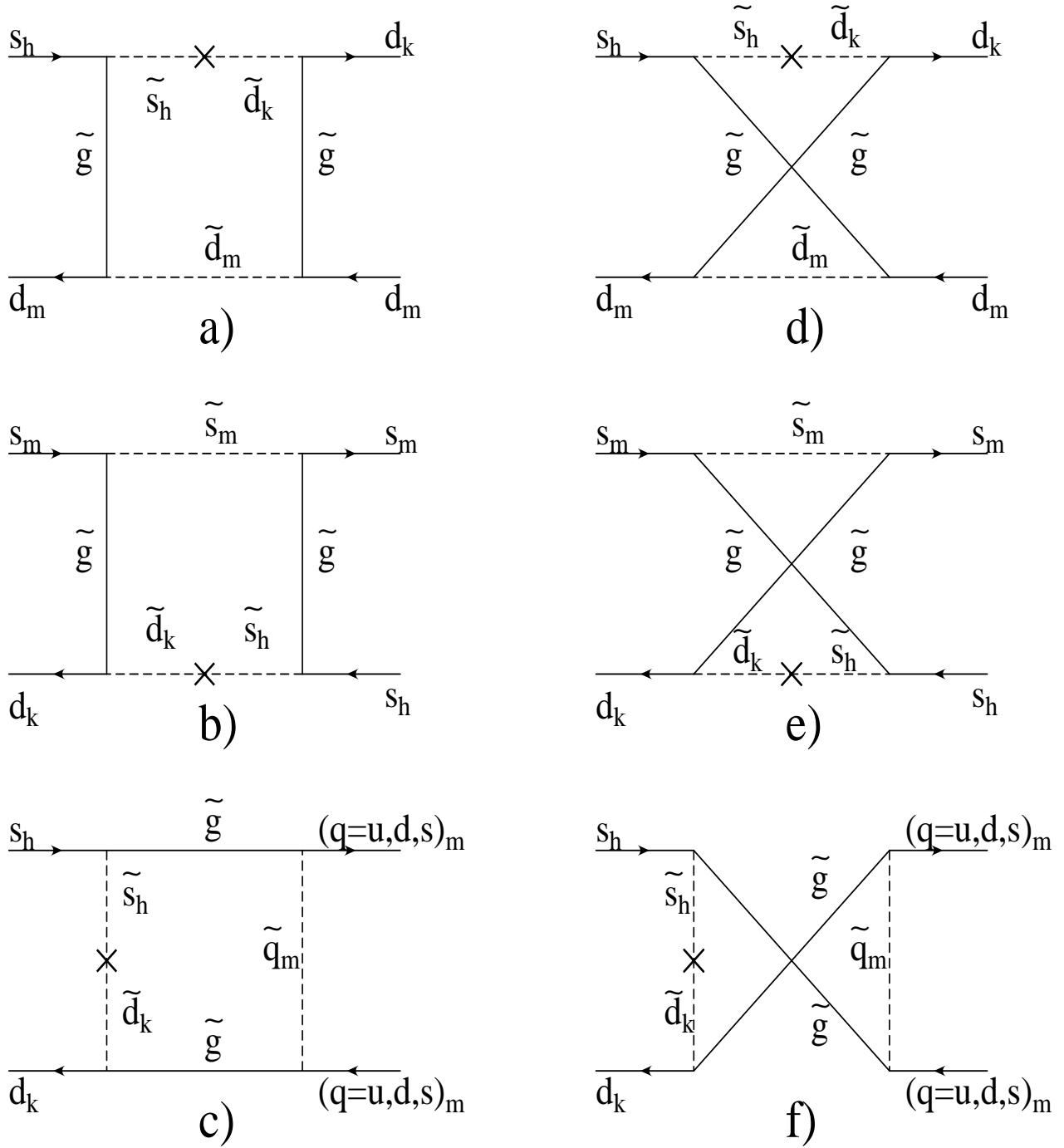} 
\caption{Box diagrams for $\Delta S=1$ transitions, with $h,k,m=\{L,R\}$.}
\label{figboxds1}
\end{figure}

\begin{figure}     
\epsfysize=19.0cm 
\epsfxsize=18.0cm 
\epsffile{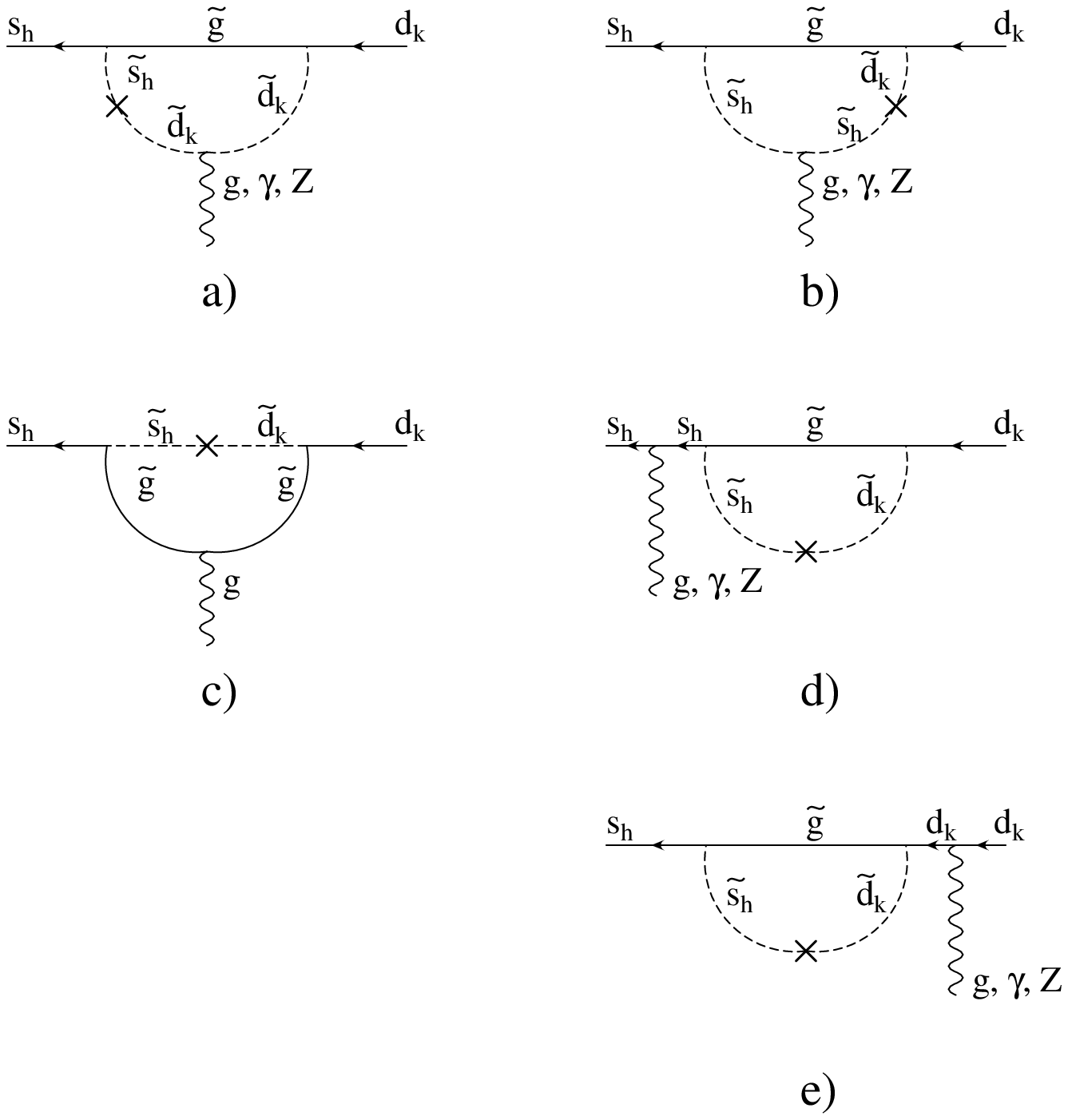} 
\caption{Penguin diagrams for $\Delta S=1$ transitions, with $h,k=\{L,R\}$.}
\label{figpeng}
\end{figure}

In the full theory, there are two classes of diagrams contributing to 
$\Delta F=1$ transitions: box diagrams (see fig.~\ref{figboxds1}) and 
penguin diagrams (see fig.~\ref{figpeng}).
In the case of gluino-mediated processes, these two contributions are of 
the same order and therefore must be included in the analysis.
While there exist in the literature various calculations of the penguin 
contribution \cite{superp}, there was no calculation of the box diagrams in 
fig.~\ref{figboxds1} before our analysis in
ref.~\cite{noi}.
A complete basis for the $\Delta S =1$ EH is the following:
\begin{eqnarray}
O_{ 3} &=& ({\bar d}^{\alpha}_{L}\gamma^{\mu} s^{\alpha}_{L})
    \sum_{q=u,d,s}({\bar q}^{\beta}_{L}\gamma_{\mu}q^{\beta}_{L}) \;,
\nonumber \\
O_{ 4} &=& ({\bar d}^{\alpha}_{L}\gamma^{\mu} s^{\beta}_{L})
    \sum_{q=u,d,s}({\bar q}^{\beta}_{L}\gamma_{\mu}q^{\alpha}_{L})\;,
\nonumber \\
O_{ 5} &=& ({\bar d}^{\alpha}_{L}\gamma^{\mu} s^{\alpha}_{L})
    \sum_{q=u,d,s}({\bar q}^{\beta}_{R}\gamma_{\mu}q^{\beta}_{R})\;,
\nonumber \\
O_{ 6} &=& ({\bar d}^{\alpha}_{L}\gamma^{\mu} s^{\beta}_{L})
    \sum_{q=u,d,s}({\bar q}^{\beta}_{R}\gamma_{\mu}q^{\alpha}_{R})\;,
\nonumber \\
O_{ 7} &=& \frac{Q_{d}e}{8\pi^2}m_{s}{\bar d}^{\alpha}_{L}\sigma^{\mu\nu}
           s^{\alpha}_{R}F_{\mu\nu}\;,
\nonumber \\
O_{ 8} &=& \frac{g}{8\pi^2}m_{s}{\bar d}^{\alpha}_{L}\sigma^{\mu\nu}
           t^A_{\alpha\beta}s^{\beta}_{R}G^A_{\mu\nu}\;,
\label{basis}
\end{eqnarray}
plus the operators $\tilde{O}_i$  obtained from $O_i$ by the exchange 
$L \leftrightarrow R$.
Here $\sigma^{\mu\nu}=\frac{i}{2}
[\gamma^{\mu},\gamma^{\nu}]$, $\alpha$ and $\beta$ are colour
indices, $g$ and $e$ are the strong and electromagnetic couplings, 
$Q_{d}=-\frac{1}{3}$ and $m_{s}$ is the mass of the strange quark.
Evaluating the diagrams in figures~\ref{figboxds1} and \ref{figpeng}, and 
performing the matching, we obtain the Wilson coefficients,  
which are given by:
\begin{eqnarray}
C_3 &=& \frac{\alpha_s^2}{m_{\tilde{q}}^2}\left(\delta^d_{12}\right)_{LL} 
\left(  - \frac{1}{9}
{\mbox B}_1(x) - \frac{5}{9}{\mbox B}_2(x) - \frac{1}{18}{\mbox P}_1(x) - 
\frac{1}{2}{\mbox P}_2(x)\right)\;,
\nonumber  \\
C_4 &=& \frac{\alpha_s^2}{m_{\tilde{q}}^2}\left(\delta^d_{12}\right)_{LL} 
\left(  - \frac{7}{3}
{\mbox B}_1(x) + \frac{1}{3}{\mbox B}_2(x) + \frac{1}{6}{\mbox P}_1(x) + 
\frac{3}{2}{\mbox P}_2(x)\right)\;,
\nonumber  \\
C_5 &=& \frac{\alpha_s^2}{m_{\tilde{q}}^2}\left(\delta^d_{12}\right)_{LL} 
\left(   \frac{10}{9}
{\mbox B}_1(x) + \frac{1}{18}{\mbox B}_2(x) - \frac{1}{18}{\mbox P}_1(x) -
\frac{1}{2}{\mbox P}_2(x)\right)\;,
\nonumber  \\
C_6 &=& \frac{\alpha_s^2}{m_{\tilde{q}}^2}\left(\delta^d_{12}\right)_{LL} 
\left(  - \frac{2}{3}
{\mbox B}_1(x) + \frac{7}{6}{\mbox B}_2(x) + \frac{1}{6}{\mbox P}_1(x) +
\frac{3}{2}{\mbox P}_2(x)\right)\;,
\nonumber  \\
C_7 &=& \frac{\alpha_s \pi}{m_{\tilde{q}}^2}\left[\left(\delta^d_{12}
\right)_{LL} \, \, 
\frac{8}{3} {\mbox M}_3(x) + (\delta^d_{12})_{LR} \, 
\, \frac{m_{\tilde{g}}}{m_{s}} \frac{8}{3}{\mbox M}_1(x) \right]\;,
\nonumber \\
C_8 &=& \frac{\alpha_s \pi}{m_{\tilde{q}}^2}\left[\left(\delta^d_{12}
\right)_{LL} \, \, 
\left(  - \frac{1}{3} {\mbox M}_3(x) - 3{\mbox M}_4(x) \right)\right. 
\nonumber \\ 
&&+\left.\left(\delta^d_{12}\right)_{LR} \, 
\, \frac{m_{\tilde{g}}}{m_{s}} \left( - \frac{1}{3}{\mbox M}_1(x) -
3{\mbox M}_2(x)\right) \right]\;,
\label{coeff}
\end{eqnarray}
where again the coefficients $\tilde{C}_i$ can be obtained from the $C_i$
just by the exchange $L \leftrightarrow R$.
\\
The functions $B_i(x)$ which result from the calculation of the box diagrams
are given by:
\begin{eqnarray}
{\mbox B}_{1}(x) &=& \frac{1 + 4x - 5x^2 + 4x\ln (x) + 2x^2\ln (x)}
{8(1 - x)^4}\;,
\nonumber \\   
{\mbox B}_{2}(x) &=& x\: \frac{5 - 4x - x^2 + 2\ln(x) + 4x\ln(x)}
{2(1 - x)^4}   \;,   
\label{bfunc}
\end{eqnarray}
while the functions $P_{i}$ and $M_{i}$, obtained from the gluino penguins, 
are:
\begin{eqnarray}
{\mbox P}_{1}(x) &=& {{1 - 6\,x + 18\,{x^2} - 10\,{x^3} - 3\,{x^4} + 
      12\,{x^3}\,\ln (x)}\over {18\,{{\left( x-1 \right) }^5}}}\;,
\nonumber \\ 
{\mbox P}_{2}(x) &=& {{7 - 18\,x + 9\,{x^2} + 2\,{x^3} + 3\,\ln (x) - 
      9\,{x^2}\,\ln (x)}\over {9\,{{\left( x-1 \right) }^5}}}\;,
 \nonumber \\
{\mbox M}_{1} (x)&=& 4 \, {\mbox B}_{1}(x) \;,
 \nonumber \\
{\mbox M}_{2} (x)&=& - x \, {\mbox B}_{2}(x) \;,
 \nonumber \\
{\mbox M}_{3} (x)&=& {{-1 + 9\,x + 9\,{x^2} - 17\,{x^3} + 18\,{x^2}\,\ln (x) + 
      6\,{x^3}\,\ln (x)}\over {12\,{{\left( x-1 \right) }^5}}}\;,
 \nonumber \\
{\mbox M}_{4} (x)&=& {{-1 - 9\,x + 9\,{x^2} + {x^3} - 6\,x\,\ln (x) - 
      6\,{x^2}\,\ln (x)}\over {6\,{{\left( x-1 \right) }^5}}}\;.
\end{eqnarray}

Our results for the gluino penguins coincide with the results of 
ref.~\cite{hagelin}.
\subsection{Hadronic Matrix Elements}
We report here for completeness the relevant matrix elements of operators 
$O_{3-7}$, which can be found in ref.~\cite{buras}:
\begin{eqnarray}
\langle(\pi \pi)_{I=0}\vert O_3  \vert K \rangle &=& \frac{X}{12} \; , 
\nonumber  \\
\langle(\pi \pi)_{I=0}\vert O_4  \vert K \rangle &=& \frac{X}{4} \; ,
\nonumber  \\
\langle(\pi \pi)_{I=0}\vert O_5  \vert K \rangle &=& -\frac{Z}{12}\; ,
\nonumber  \\
\langle(\pi \pi)_{I=0}\vert O_6  \vert K \rangle &=& -\frac{Z}{4} \; ,
\nonumber  \\
\langle(\pi \pi)_{I=0}\vert O_8  \vert K \rangle &=& -\frac{1}{16 \pi^{2}} 
\frac{11}{2} \frac{f_{K}^{2}}{f_{\pi}^{3}} m_{K}^{2}m_{\pi}^{2} \; ,
\nonumber  \\
\langle(\pi \pi)_{I=2}\vert O_i  \vert K \rangle &=& 0 \;\;\;\;
\;\;\;\; i=3, \ldots ,6\; ,
\label{meds1}
\end{eqnarray}
where
\begin{eqnarray}
X&=&f_{\pi}\,\left(m_{K}^{2}-m_{\pi}^{2}\right) \; , \nonumber  \\
Z&=&4 \, \left(\frac{f_{K}}{f_{\pi}}-1\right)f_{\pi}\left(
\frac{m_{K}^{2}}{m_{s}+m_{d}}\right)^{2} \; .
\end{eqnarray}
The matrix elements of the operators $\tilde{O}_i$ can be obtained from the 
matrix elements of $O_i$ multiplying them by $(-1)$.
\subsection{General Analysis of $\Delta F=1$ Processes}
We now present the results of a model-independent analysis of $\Delta 
F=1$ processes. We will start with $\Delta S=1$ transitions, and in 
particular with the CP-violating parameter $\varepsilon^{\prime}/\varepsilon$, 
then we proceed to $\Delta B=1$ and consider radiative B decays, and continue with 
the constraints coming from radiative decays in the 
leptonic sector: $\mu \to e \gamma$, $\tau \to \mu \gamma$ and 
$\tau \to e \gamma$. We will close this Section with limits on 
flavour-diagonal Left-Right mass terms coming from electric dipole moments and 
from radiative corrections to masses.

The expression for $\varepsilon^{\prime}$ is the following (see 
for example ref.~\cite{buras}):
\begin{equation}
\varepsilon^{\prime} = i \frac{e^{i(\delta_{2}-\delta_{0})}}{\sqrt{2}}
\frac{\omega}{\Re A_{0}}\left( \omega^{-1}\Im A_{2} - \Im A_{0}\right) \; ,
\label{epp}
\end{equation}
where $\omega=\Re A_2/\Re A_0$, and the amplitudes are defined as:
\begin{equation}
A_Ie^{i\delta_I}=\langle \pi \pi(I)\vert \cal{H}_{eff}\vert K^{0}\rangle,
\end{equation}
where $I=0,2$ is the isospin of the final two-pion state
and the $\delta_I$'s are the strong phases induced by final-state interaction.

\begin{figure}   
    \begin{center}
    \epsfysize=8truecm
    \leavevmode\epsffile{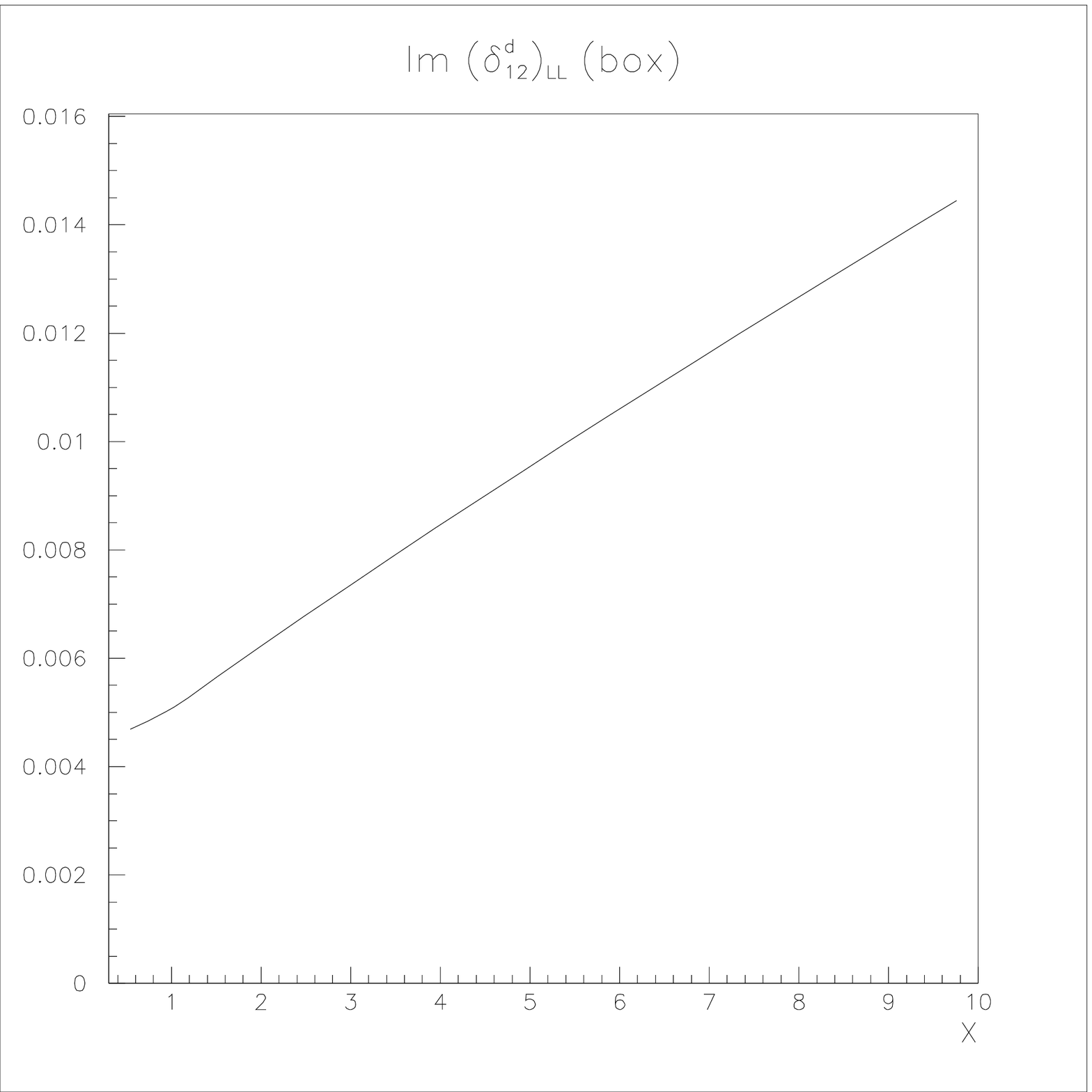}
    \end{center}
    \caption[]{The $\mbox{Im}\left(\delta^d_{12}\right)_{LL}$ as a function
     of $x=m_{\tilde{g}}^2/m_{\tilde{q}}^2$, obtained considering only the 
     box contribution, for  an average squark mass 
     $m_{\tilde{q}}=100\mbox{GeV}$.}
     \label{eppllbox}
\end{figure}

\begin{figure}   
    \begin{center}
    \epsfysize=8truecm
    \leavevmode\epsffile{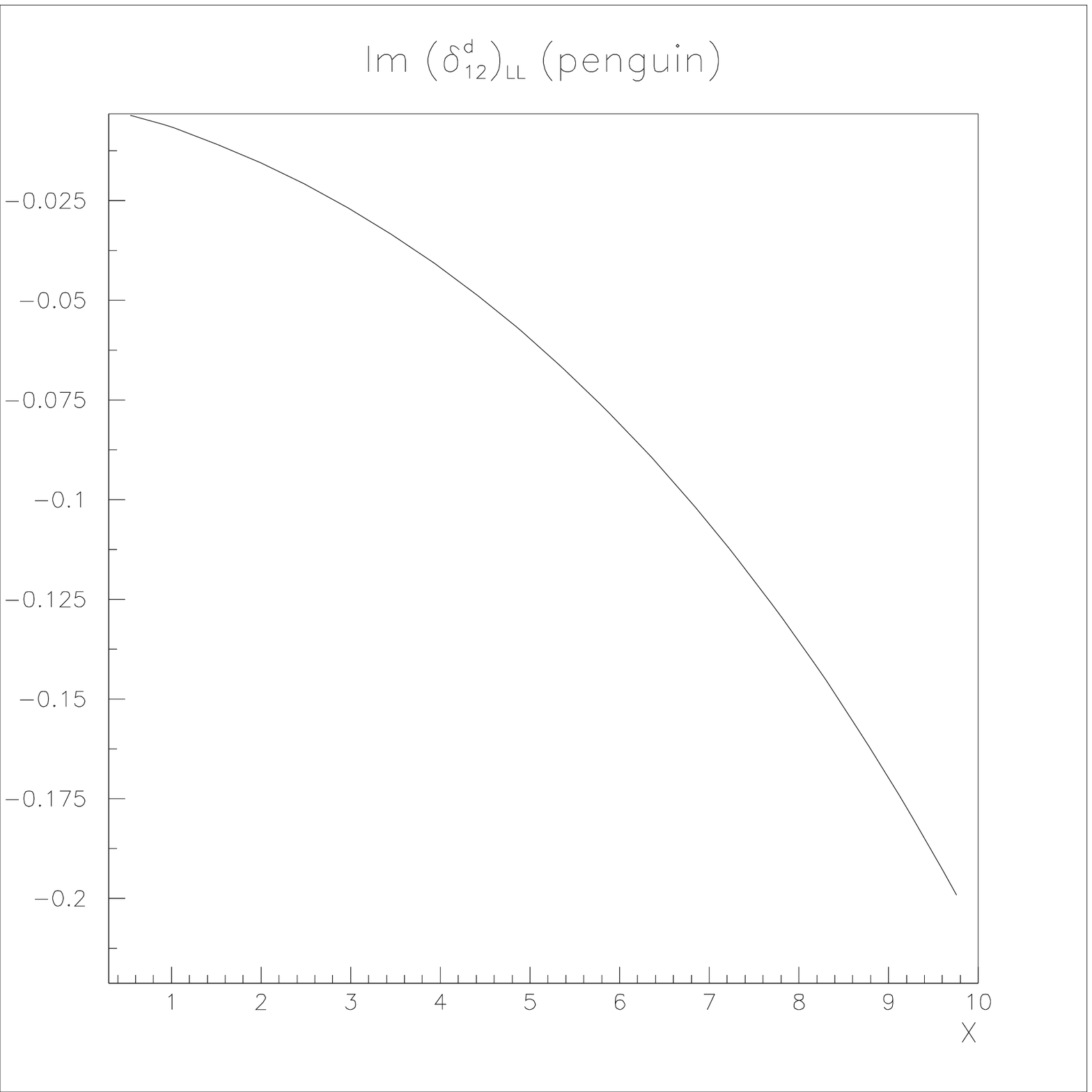}
    \end{center}
    \caption[]{The $\mbox{Im}\left(\delta^d_{12}\right)_{LL}$ as a function
     of $x=m_{\tilde{g}}^2/m_{\tilde{q}}^2$, obtained considering only the 
     penguin contribution, for  an average squark mass 
     $m_{\tilde{q}}=100\mbox{GeV}$.}
     \label{eppllpeng}
\end{figure}

\begin{figure}   
    \begin{center}
    \epsfysize=8truecm
    \leavevmode\epsffile{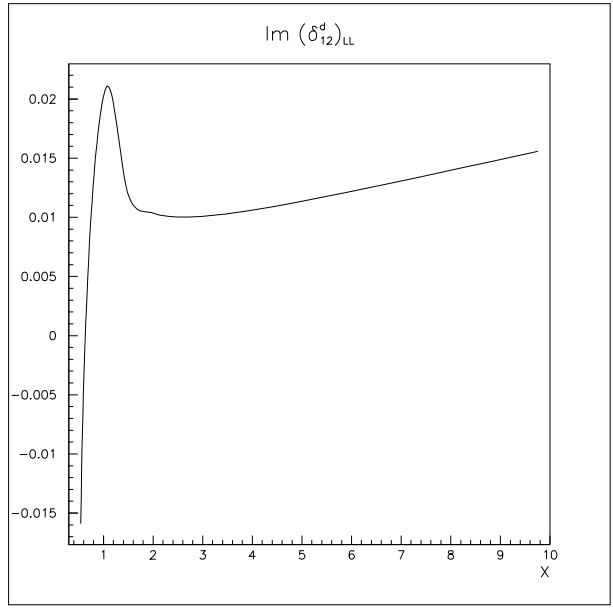}
    \end{center}
    \caption[]{The $\mbox{Im}\left(\delta^d_{12}\right)_{LL}$ as a function
     of $x=m_{\tilde{g}}^2/m_{\tilde{q}}^2$, for  an average squark mass 
     $m_{\tilde{q}}=100\mbox{GeV}$.}
     \label{eppll}
\end{figure}

Imposing that the supersymmetric contribution to $\varepsilon^{\prime}/
\varepsilon$
 does not 
exceed in absolute value the upper limit in table~\ref{ul}, we obtain the 
conservative limits given in table~\ref{imds1}. 
It is interesting to note that the contributions of box and penguin 
diagrams to the $LL$ terms in (\ref{epp}), 
which are separately plotted in figs.~\ref{eppllbox}  and \ref{eppllpeng} 
respectively, have opposite signs and therefore tend to 
cancel each other in the region around $x=1$, where they are of the same 
size. This is the reason of the peak around $x=1$ in the plot of the 
complete contribution, given in fig.~\ref{eppll}.

A comment is necessary at this point. In the SM, the contribution to 
$\varepsilon^\prime$ coming from  
electropenguin operators is non-negligible for a heavy top. 
One might therefore wonder whether 
this remains true in SUSY or not. In particular, we have considered the 
contribution to $\varepsilon^\prime$ coming from gluino-mediated electroweak 
penguins. \\
Let us first note that gluino-mediated $Z^0$-penguins are suppressed 
by a factor of $m_s/M_Z$, relative to $\gamma$-penguins \cite{bsg}. 
This is due to the 
fact that the effective $b-s-Z$ vertex is proportional, in the case of 
gluino-mediated transitions, to the effective $b-s-\gamma$ vertex, apart 
from chirality breaking terms of order $m_s$. Now, for gauge invariance, the 
effective $b-s-\gamma$ vertex must be proportional either to $(\gamma_\mu q^2 - 
q_\mu \qslash )$ or to $m_s \sigma_{\mu\nu}q^\nu$. 
The first form factor is the one 
which originates the so-called electropenguin operators. In the case of the 
photon, the $q^2$ factor cancels the pole of the propagator, giving an 
effective four-fermion operator, while in the case of the $Z$ boson we get a 
suppression of order $q^2/M_Z^2$ which makes these contributions negligible. 
We are thus left with the chirality breaking terms, which are themselves of 
order $m_s/M_Z$, and therefore can be safely neglected. \\
The contribution to $\varepsilon^\prime$ coming from gluino-mediated 
$\gamma$-superpenguins 
has  got no explicit suppression factor: however, we have numerically 
checked that this contribution is negligible.
 \begin{table}
 \begin{center}
 \begin{tabular}{||c|c|c||}  \hline \hline
  & & \\
  $x$ & ${\scriptstyle\left|\Im \left(\delta^{d}_{12}  \right)_{LL}
\right|} $ &
 ${\scriptstyle\left|\Im \left(\delta^{d}_{12}\right)  _{LR}\right| }$ 
\\
  & & \\\hline
 $
   0.3
 $ &
 $
1.0\times 10^{-1}
 $ & $
1.1\times 10^{-5}
 $ \\
 $
   1.0
 $ &
 $
4.8\times 10^{-1}
 $ & $
2.0\times 10^{-5}
 $ \\
 $
   4.0
 $ &
 $
2.6\times 10^{-1}
 $ & $
6.3\times 10^{-5}
 $ \\ \hline \hline
 \end{tabular}
 \caption[]{Limits from $\varepsilon^{\prime}/\varepsilon < 2.7 \times 10^{-3}$ 
 on  $\Im\left(\delta_{12}^{d}\right)$, for 
 an average squark mass $m_{\tilde{q}}=500\mbox{GeV}$ and for different values of 
 $x=m_{\tilde{g}}^2/m_{\tilde{q}}^2$. For different values of $m_{\tilde{q}}$, 
 the limits can be obtained multiplying the ones in the table by 
 $\left(m_{\tilde{q}}(\mbox{GeV})/500\right)^2$.}
 \label{imds1}
 \end{center}
 \end{table}

We now consider the decay $b \to s \gamma$. The gluino-mediated 
contribution is given by (see ref.~\cite{gabbmas}):
\begin{eqnarray}
	{\mbox{BR}}(b \to s \gamma) &=& \frac{\als^{2}\alpha}{81 \pi^{2} 
	m_{\tilde{q}}^{4}}m_{b}^{3}\tau_{B} \Biggl\{\biggl\vert m_{b} M_{3}(x) 
	\left(\delta^{d}_{23}\right)_{LL}
	\nonumber \\ 
	&&+ m_{\tilde{g}} M_{1}(x) 
	\left(\delta^{d}_{23}\right)_{LR}\biggr\vert^{2} + L 
	\leftrightarrow R \Biggr\}\; .
	\label{brbsg}
\end{eqnarray}
Using the result (\ref{brbsg}) we can obtain the limits in table~\ref{tab:bsg}, 
by imposing that each individual term in eq.~(\ref{brbsg}) does not exceed in 
absolute value the range given in tab.~\ref{ul}, which includes the uncertainty
coming from the SM prediction.
Table~\ref{tab:bsg} shows that the decay
$(b\rightarrow s+\gamma)$ does not limit the $(\delta^d_{23})_{LL}$ insertion
for a SUSY breaking of O(500 GeV). Indeed, even taking
$m_{\tilde{q}}=100\mbox{GeV}$, the term $(\delta^d_{23})_{LL}$ is 
only marginally
limited ( $(\delta_{23}^d)_{LL}<0.3$ for $x=1$). 
Obviously, $(\delta_{23}^d)_{LR}$ 
is much more constrained since with a $\delta_{LR}$ FC mass insertion the 
helicity flip needed for $(b\rightarrow s+\gamma)$ is realized in the gluino
internal line and so this contribution has an amplitude enhancement
of a factor $m_{\tilde{g}}/m_b$ over the previous case with 
$\delta_{LL}$.
 \begin{table}
 \begin{center}
 \begin{tabular}{||c|c|c||}  \hline \hline
  & & \\
 $x$ & $\left|\left(\delta^{d}_{23} \right)_{LL}\right| $ &
 $\left|  \left(\delta^{d}_{23} \right)_{LR}\right| $ \\
  & & \\ \hline
 $
   0.3
 $ &
 $
4.4
 $ & $
1.3\times 10^{-2}
 $ \\
 $
   1.0
 $ &
 $
8.2
 $ & $
1.6\times 10^{-2}
 $ \\
 $
   4.0
 $ &
 $
26
 $ & $
3.0\times 10^{-2}
 $ \\ \hline \hline
 \end{tabular}
 \caption[]{Limits on the $\left| \delta_{23}^{d}\right|$ from
 $b\rightarrow s \gamma$ decay for an average squark mass $m_{\tilde{q}}=500\mbox{GeV}$ 
 and for different values of $x=m_{\tilde{g}}^2/m_{\tilde{q}}^2$. For different values of $m_{\tilde{q}}$, 
 the limits can be obtained multiplying the ones in the table by 
 $\left(m_{\tilde{q}}(\mbox{GeV})/500\right)^2$.}
 \label{tab:bsg}
 \end{center}
 \end{table}

\begin{figure}   
    \begin{center}
    \epsfysize=8truecm
    \leavevmode\epsffile{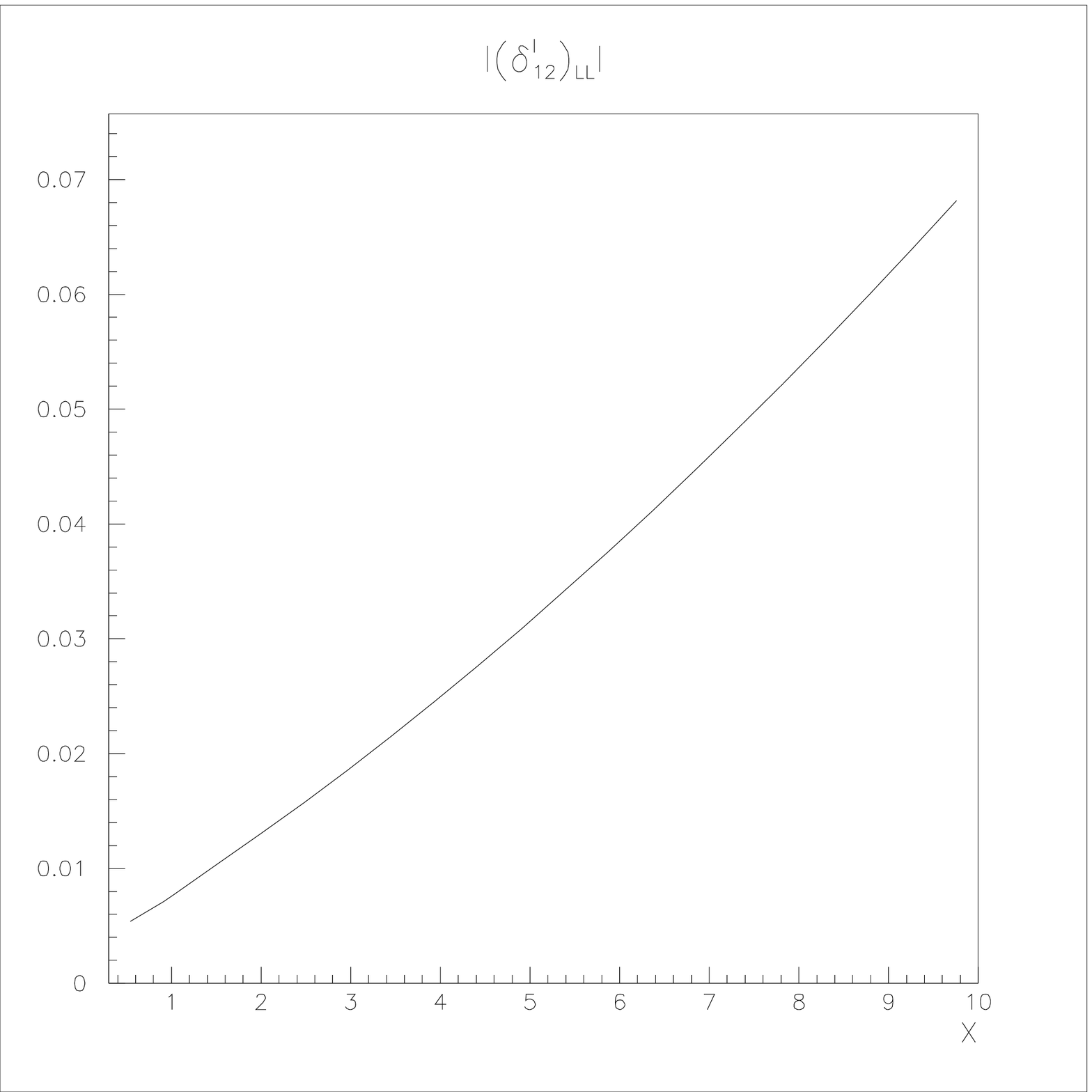}
    \end{center}
    \caption[]{The $\vert\left(\delta^l_{12}\right)_{LL}\vert$ as a function
     of $x=m_{\tilde{\gamma}}^2/m_{\tilde{l}}^2$, for  an average slepton mass 
     $m_{\tilde{l}}=100\mbox{GeV}$.}
    \protect\label{muegll}
\end{figure}

\begin{figure}   
    \begin{center}
    \epsfysize=8truecm
    \leavevmode\epsffile{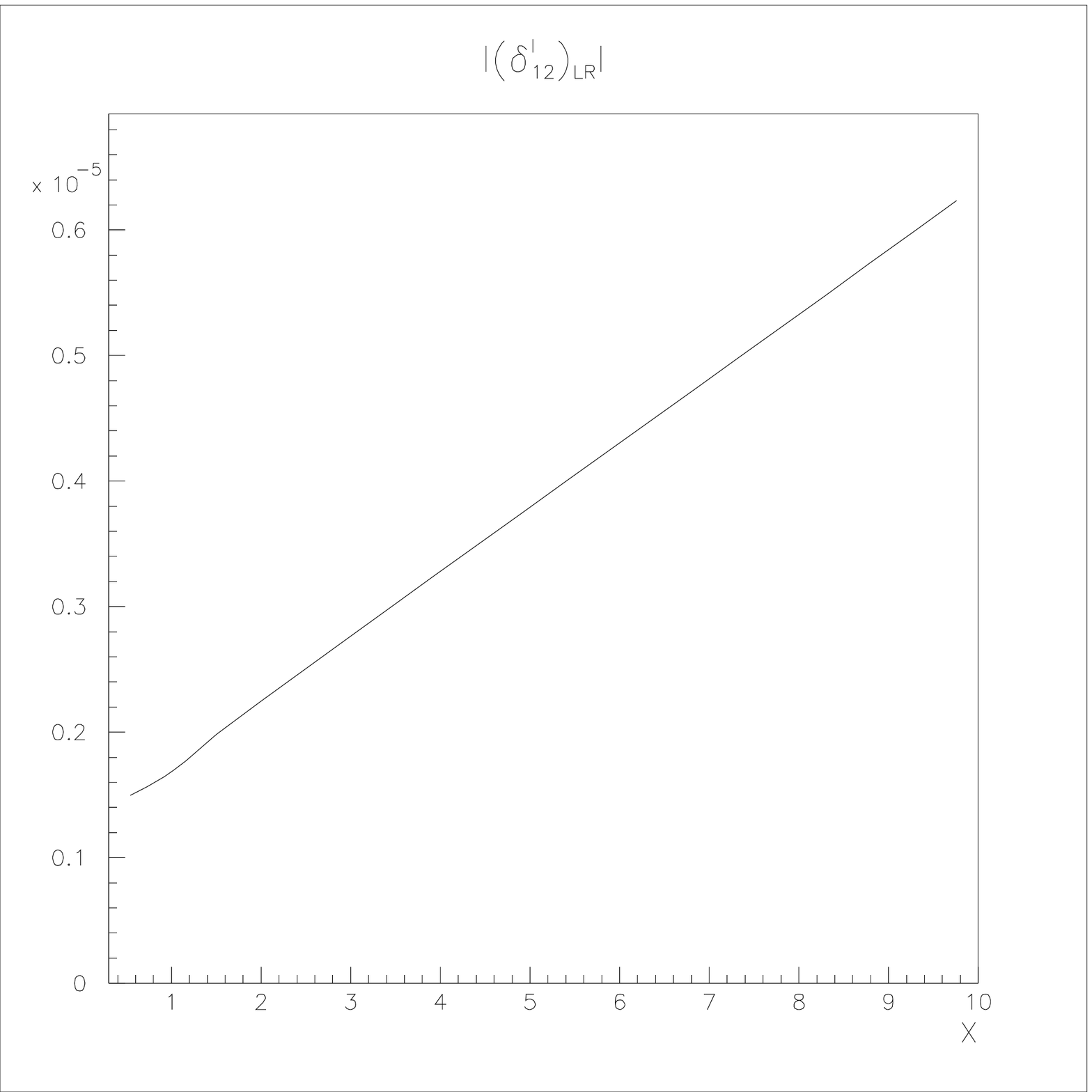}
    \end{center}
    \caption[]{The $\vert\left(\delta^l_{12}\right)_{LR}\vert$ as a function
     of $x=m_{\tilde{\gamma}}^2/m_{\tilde{l}}^2$, for  an average slepton mass 
     $m_{\tilde{l}}=100\mbox{GeV}$.}
    \protect\label{mueglr}
\end{figure}

A similar analysis can be performed in the leptonic sector where the masses
$m_{\tilde{q}}$ and $m_{\tilde{g}}$ are replaced by the average slepton
mass $m_{\tilde{l}}$ and the photino mass $m_{\tilde{\gamma}}$ respectively.
The branching ratio for the process $l_{i} \to l_{j} \gamma$ is the following 
(see ref.~\cite{gabbmas}):
\begin{eqnarray}
	{\mbox{BR}}(l_{i} \to l_{j} \gamma) &=& \frac{\alpha^{3}}{G_{F}^{2}}
	\frac{12 \pi} 
	{m_{\tilde{l}}^{4}}\Biggl\{\biggl\vert M_{3}(x) 
	\left(\delta^{l}_{ij}\right)_{LL} 
	\nonumber \\
	&&+ \frac{m_{\tilde{\gamma}}}{m_l} 
	M_{1}(x) \left(\delta^{l}_{ij}\right)_{LR}\biggr\vert^{2}
	+ L \leftrightarrow R \Biggr\} \cdot 
	{\mbox{BR}}(l_{i} \to l_{j} \nu_{i} \bar{\nu}_{j}) \; .
	\label{litolj}
\end{eqnarray}
In table~\ref{lep} we exhibit the bounds on $\left(\delta^l_{ij}\right)_{LL}$ 
and $\left(\delta^l_{ij}\right)_{LR}$ coming from the limits on 
$\mu\rightarrow e\gamma,~\tau\rightarrow e\gamma$ and 
$\tau\rightarrow \mu\gamma$, for a slepton mass of O(100 GeV)
and for different values of $x=m_{\tilde{\gamma}}^2/m_{\tilde{l}}^2$.
The dependence of those limits on $x$ is given in fig.~\ref{muegll} and 
\ref{mueglr}.
Our results confirm those obtained in refs \cite{gabbmas, hagelin}.
 \begin{table}
 \begin{center}
 \begin{tabular}{||c|c|c||}  \hline \hline
  & & \\ 
 $x$ & $\left|\left(\delta^{l}_{12} \right)_{LL}\right| $ &
 $\left|  \left(\delta^{l}_{12} \right)_{LR}\right| $ \\ 
  & & \\ \hline
 $
   0.3
 $ &
 $
4.1\times 10^{-3}
 $ & $
1.4\times 10^{-6}
 $ \\
 $
   1.0
 $ &
 $
7.7\times 10^{-3}
 $ & $
1.7\times 10^{-6}
 $ \\
 $
   5.0
 $ &
 $
3.2\times 10^{-2}
 $ & $
3.8\times 10^{-6}
 $ \\ \hline \hline
  & & \\ 
 $x$ & $\left|\left(\delta^{l}_{13} \right)_{LL}\right| $ &
 $\left|  \left(\delta^{l}_{13} \right)_{LR}\right| $ \\ 
  & & \\ \hline
 $
   0.3
 $ &
 $
15
 $ & $
8.9\times 10^{-2}
 $ \\
 $
   1.0
 $ &
 $
29
 $ & $
1.1\times 10^{-1}
 $ \\
 $
   5.0
 $ &
 $
1.2\times 10^{2}
 $ & $
2.4\times 10^{-1}
 $ \\ \hline \hline
  & & \\ 
 $x$ & $\left|\left(\delta^{l}_{23} \right)_{LL}\right| $ &
 $\left|  \left(\delta^{l}_{23} \right)_{LR}\right| $ \\
  & & \\ \hline
 $
   0.3
 $ &
 $
2.8
 $ & $
1.7\times 10^{-2}
 $ \\
 $
   1.0
 $ &
 $
5.3
 $ & $
2.0\times 10^{-2}
 $ \\
 $
   5.0
 $ &
 $
22
 $ & $
4.4\times 10^{-2}
 $ \\ \hline \hline
 \end{tabular}
 \caption[]{Limits on the $\left| \delta_{ij}^{d}\right|$ from
 $l_j\rightarrow l_i \gamma$ decays for 
 an average slepton mass $m_{\tilde{l}}=100\mbox{GeV}$ and for different values of 
 $x=m_{\tilde{\gamma}}^2/m_{\tilde{l}}^2$. 
For different values of $m_{\tilde{l}}$, 
 the limits can be obtained multiplying the ones in the table by 
 $\left(m_{\tilde{l}}(\mbox{GeV})/100\right)^2$.}
 \label{lep}
 \end{center}
 \end{table}

We conclude this Section with the analysis of the limits on flavour-conserving
$\left(\delta_{ii}\right)_{LR}$ mass insertions coming from 
radiative corrections to mass terms and electric dipole 
moments\footnote{We thank R. Barbieri for suggesting this point to us.}.
 \begin{table}
 \begin{center}
 \begin{tabular}{||c|c|c|c||}  \hline \hline
  & & & \\
 $x$ & $\left|\Re\left(\delta^{d}_{11} \right)_{LR}\right| $ &
 $\left| \Re \left(\delta^{d}_{22} \right)_{LR}\right| $ &
 $\left| \Re \left(\delta^{d}_{33} \right)_{LR}\right| $ \\
  & & & \\ \hline
 $
   0.3
 $ &
 $
2.1\times 10^{-3}
 $ & $
3.1\times 10^{-2}
 $ & $
9.6\times 10^{-1}
 $ \\
 $
   1.0
 $ &
 $
1.6\times 10^{-3}
 $ & $
2.4\times 10^{-2}
 $ & $
7.3\times 10^{-1}
 $ \\
 $
   4.0
 $ &
 $
1.4\times 10^{-3}
 $ & $
2.1\times 10^{-2}
 $ & $
6.5\times 10^{-1}
 $ \\ \hline \hline
  & & & \\
 $x$ & $\left|\Re \left(\delta^{l}_{11} \right)_{LR}\right| $ &
 $\left|\Re  \left(\delta^{l}_{22} \right)_{LR}\right| $ &
 $\left|\Re  \left(\delta^{l}_{33} \right)_{LR}\right| $ \\
  & & & \\ \hline
 $
   0.3
 $ &
 $
1.1\times 10^{-2}
 $ & $
2.1
 $ & $
36
 $ \\
 $
   1.0
 $ &
 $
8.0\times 10^{-3}
 $ & $
1.6
 $ & $
27
 $ \\
 $
   5.0
 $ &
 $
7.1\times 10^{-3}
 $ & $
1.4
 $ & $
24
 $ \\ \hline \hline
 \end{tabular}
 \caption[]{Limits on $\Re \left(\delta_{ii} \right)_{LR}$ from one-loop mass 
terms, for $m_{\tilde{q}}=500$ GeV and $m_{\tilde{l}}=100$ GeV. }
 \label{masses}
 \end{center}
 \end{table}
 \begin{table}
 \begin{center}
 \begin{tabular}{||c|c|c|c||}  \hline \hline
  & & & \\
 $x$ & $\left|\Im\left(\delta^{d}_{11} \right)_{LR} \right|$ &
 $\left|\Im  \left(\delta^{u}_{11} \right)_{LR} \right|$ &
 $\left|\Im  \left(\delta^{l}_{11} \right)_{LR} \right|$\\
  & & & \\ \hline
 $
   0.3
 $ &
 $
2.4\times 10^{-6}
 $ & $
4.9\times 10^{-6}
 $ & $
3.0\times 10^{-7}
 $ \\
 $
   1.0
 $ &
 $
3.0\times 10^{-6}
 $ & $
5.9\times 10^{-6}
 $ & $
3.7\times 10^{-7}
 $ \\
 $
   4.0
 $ &
 $
5.6\times 10^{-6}
 $ & $
1.1\times 10^{-5}
 $ & $
7.0\times 10^{-7}
 $ \\ \hline \hline
 \end{tabular}
 \caption[]{Limits on $\Im \left(\delta_{ii} \right)_{LR}$ from electric 
dipole moments, for $m_{\tilde{q}}=500$ GeV and $m_{\tilde{l}}=100$ GeV.}
  \label{dipoles}
 \end{center}
 \end{table}

A Left-Right diagonal mass insertion $(\delta_{ii})_{LR}=
(\delta_{ii})_{RL}$ generates a one-loop mass term given by
\begin{equation}
\Delta m_i=-\frac{4}{3}\frac{2 \als}{4 \pi} 
m_{\tilde{g}} \Re \left(\delta^q_{ii}\right)_{LR}\,
I(x) \;  
\label{dmq}
\end{equation}
for quarks, and by
\begin{equation}
\Delta m_i=-\frac{2 \alpha}{4 \pi} m_{\tilde{\gamma}} 
\Re \left(\delta^l_{ii}\right)_{LR}\,
I(x) \;  
\label{dml}
\end{equation}
for leptons. The function $I(x)$ is given by
\begin{equation}
I(x)=\frac{-1+x-x \ln(x)}{(1-x)^2} \; .
\end{equation} 
Imposing that these mass terms do not exceed in absolute 
value the masses of 
quarks and leptons (for the $d$ quark we have used a value of 10 
MeV), we obtain the limits given in table~\ref{masses}.

Limits on the imaginary parts of $\left(\delta_{ii}\right)_{LR}$
can be obtained by analyzing the contribution to the Electric Dipole Moments 
(EDM) of the neutron and of the electron. The EDM of the electron and of $u$ 
and $d$ quarks are given by:
\begin{eqnarray}
\frac{d_e}{e}&=&-\frac{\alpha m_{\tilde{\gamma}}}{2 
\pi m_{\tilde{l}}^2}\, M_{1}(x) \,
\Im \left(\delta^l_{11}\right)_{LR} \; , \nn \\
\frac{d_d}{e}&=&-\frac{2}{9}
\frac{\als m_{\tilde{g}}}{\pi m_{\tilde{q}}^2}\, M_{1}(x) \,
\Im \left(\delta^d_{11}\right)_{LR} \; , \nn \\
\frac{d_u}{e}&=&\frac{4}{9}
\frac{\als m_{\tilde{g}}}{\pi m_{\tilde{q}}^2}\, M_{1}(x) \,
\Im \left(\delta^u_{11}\right)_{LR} \; 
\end{eqnarray}
and in the quark model the EDM of the neutron is given by 
\begin{equation}
d_n=\frac{1}{3}\left(4 d_d - d_u\right)\; .
\end{equation}
Imposing that each of the above contributions does not exceed in absolute 
value the limits in table~\ref{ul}, we obtain the upper bounds in 
table~\ref{dipoles}. 

The limits obtained from mass terms and EDM's are limits on diagonal terms, 
not on flavour-violating ones. However, in general one expects that 
off-diagonal terms should be proportional to diagonal ones via some (small) 
mixing angle, and therefore the limits above might be interpreted as indirect 
upper bounds on flavour violating terms.

If one compares the constraints in tables~\ref{masses} and \ref{dipoles} with 
the bounds on the FC $\delta$'s involving the first two generations, one 
notices that only the bound on $\Im \left(\delta^d_{11}\right)_{LR}$ is more 
severe than the corresponding limit on $\Im \left(\delta^d_{12}\right)_{LR}$.
This means that if we envisage a situation where the diagonal entries are 
larger than the corresponding off-diagonal entries there is no chance to 
obtain a large contribution to $\varepsilon^\prime/\varepsilon$ from SUSY 
while respecting the constraint from $d_{n}$.
\section{QCD corrections}
\label{sec:qcd}
In the framework of the Standard Model (SM), strong interactions are known to 
give sizeable contributions to FCNC processes, and in particular to $b 
\to s \gamma$ decays and to $\varepsilon^\prime/\varepsilon$. 
It is therefore important to know whether QCD corrections could be so 
important also in our case, and how could the inclusion of these corrections
modify the limits we previously obtained.

Let us start by considering $b \to s \gamma$ decays. In the SM the effect 
of QCD corrections is a dramatic enhancement of the rate, of a factor 
$\sim 2-5$ \cite{bsgQCD}. 
This is due to the mixing between the magnetic moment operator $O_7$ and the 
four fermion operator $O_2=
\bar{s}^{\alpha}_{L} \gamma_{\mu} c^{\alpha}_{L}\,\bar{c}^{\beta}_{L}
\gamma^{\mu} b^{\beta}_{L}$, whose coefficient at the scale $M_W$ is 
$\sim 10$ times bigger. \\
Since the mixing with $O_{8}$ is much smaller, 
we can use the following approximation for the coefficient of the 
operator $O_7$ at a scale $\mu \sim m_{b}$ \cite{bsgburas}:
\begin{equation}
C_{7}(\mu)\simeq\eta^{\frac{16}{23}}C_{7}(M_{W})+C_{2}(M_{W})\sum_{i=
1}^{8}h_{i}\eta^{\alpha_{i}}\; ,
\label{bsgcorr}
\end{equation}
where $\eta=\als(M_{W})/\als(\mu)$ and $h_{i}$ and $\alpha_{i}$ are related to 
the anomalous dimensions of the operators $O_{i}$. \\
Now we note that the last term in the r.h.s. of eq.~(\ref{bsgcorr}), which is 
responsible for the enhancement, remains exactly the same when we include 
supersymmetric contributions. Indeed, gluino-mediated transitions only affect 
$C_{7}(M_{W})$ in eq.~(\ref{bsgcorr}). Therefore, if one takes into account 
the prediction of the standard model, including QCD corrections, when 
comparing the SUSY contributions due to gluino exchange with the experimental 
value for the decay $b \to s \gamma$, there is no need to explicitly consider 
QCD corrections to the SUSY contributions. \\
Of course, as previously noticed, one should bear in mind that to calculate the full 
amplitude for $b \to s \gamma$ decays in any definite SUSY model, one should 
include all SUSY contributions (chargino and neutralino mediated) and SM 
ones, and consider possible interference effects and cancellations
\cite{bsg, bsgfull}.

Concerning $\varepsilon^\prime/\varepsilon$, one has to note two main 
points:
\begin{enumerate}
	\item  the constraints from $\varepsilon^\prime$ dominate 
	over the ones from $\varepsilon$ only in the $LR$ sector, where just 
	the magnetic moment operator contributes, and for the reasons given 
	above QCD corrections are negligible in this case;
	
	\item  in the $LL$ and $RR$ sectors the main contribution to
	$\varepsilon^\prime$ in the strong sector is given by the operator 
	$O_6$. Numerically, we find
	that the effect of leading order QCD corrections is to enhance its 
	coefficient 
	by a factor $\leq 2$. 
	This means that even with the inclusion of QCD corrections  
	the dominant constraints in the $LL$ and $RR$ sectors 
	come from $\varepsilon$. Therefore, the results of our analysis remain 
	unaffected.
\end{enumerate}
\section{FCNC in SUSY-GUT's and in SUSY with Non-Universal Soft Breaking Terms}
\label{sec:models}
In this Section we consider two classes of SUSY theories with potentially 
interesting (or dangerous, according to the viewpoint) contributions to FCNC.

The first class includes GUT extensions of the MSSM where the presence of 
quark and lepton superfields in common supermultiplets and the large value of 
the top Yukawa coupling produce very conspicuous contributions, in particular 
in the leptonic sector.

The second class comprises models where the breaking 
of SUSY gives rise to soft breaking terms which are not flavour-universal.
This is a common feature of effective supergravities that are the point-like 
limit of four-dimensional superstrings. Only with rather specific assumptions 
on the mechanism of SUSY breaking can one avoid the occurrence of such 
non-universality.

Obviously, there already exist rather exhaustive analyses of both these 
classes of SUSY theories in relation to the FCNC problem. It is not our 
purpose here to produce yet another analysis of this kind, but, rather, we 
intend to estimate the values of some $\delta$ mass insertions in a few 
examples within the two above-mentioned classes of generalized SUSY theories 
and to compare them with the bounds on the $\delta$'s that we derived in the 
previous Sections. An interesting aspect of our analysis will be the 
comparison with the ``exact'' results that have been obtained in the 
literature using the physical (mass eigenstates) basis for fermions and 
sfermions. We will discuss to what extent the mass insertion approach is valid 
and how it has to be correctly implemented to get results which are quite 
close to the full computation in the mass eigenstate basis.
\subsection{SUSY SU(5)}
\label{sec:su(5)}
It has been known since the pioneering works of Duncan and Donoghue, Nilles 
and Wyler in 1983 \cite{FCNC} 
that even in the MSSM the running of sfermion masses from 
the superlarge scale where SUSY is broken down to the Fermi scale is 
responsible for a misalignment of fermion and sfermion mass matrices with the 
consequent presence of FC in $\tilde{g}-f-\tilde{f}$ or $\tilde{\gamma}-f-
\tilde{f}$ vertices. 

The key-feature of the unification of quark and lepton 
superfields into larger multiplets in SUSY-GUT's in relation to the FCNC issue 
was thoroughly investigated by Hall, Kostelecky and Rabi ten years ago 
\cite{mins}.
But it was only recently, with the realization of the large size of the top 
Yukawa coupling, that it became clear that in SUSY-GUT's radiative corrections 
can lead to slepton non-degeneracies so important as to imply $L_e$ and $L_\mu$ 
violations just in the ballpark of the present or near future experimental 
range \cite{barbhall}.

The interested reader can find all the details of this relevant low-energy 
manifestation of grand unification in the works of 
refs.~\cite{barbieri, fcncgut}.
Here we limit ourselves to a computation of $\mbox{BR}(\mu \to e \gamma)$ in 
SUSY SU(5) using the mass insertion approximation. As  we will see some of the 
$\delta$ quantities in SU(5) turn out to be only one order of magnitude 
smaller than the corresponding upper bounds that we found in  Sect.~\ref{df1} 
when analyzing the bounds coming from $\mbox{BR}(\mu \to e \gamma)$.

We follow closely the notation of ref.~\cite{barbieri} where further details 
can be found.
Denoting with $T$ and $\bar{F}$ the $\underline{10}$ and $\underline{\bar{5}}$ 
representations of SU(5) of standard matter and with $H$ and $\bar{H}$ the 
$\underline{5}$ and $\underline{\bar{5}}$ Higgs supermultiplets, the 
superpotential reads:
\begin{equation}\label{eq:WYSU(5)}
W=T_i    \lambda ^{\rm u}_{ij} T_j H +
  T_i    \lambda ^{\rm d}_{ij}\bar{F}_j\bar{H}\equiv
  T^T\mb{\lambda}^{\rm u} T H +
  T^T\mb{\lambda}^{\rm d}\bar{F}\bar{H}\; ,
\end{equation}
from the Planck scale down to the grand unification scale $M_{\rm G}$.
The scalar soft breaking terms, above $M_{\rm G}$, are given by:
\begin{equation}\label{eq:Vsoft5}
V_{\rm soft}=
      \tilde{T}^\dagger \mb{m}_T^2\tilde{T}+
\tilde{\!\bar{F}}^\dagger \mb{m}_{\bar{F}}^2 \tilde{\!\bar{F}}+
m_H^2 |H|^2 + m_{\bar{H}}^2 |\bar{H}|^2+
\tilde{T}^T \mb{A}^{\rm u}\mb{\lambda}^{\rm u}\tilde{T} H+
\tilde{T}^T \mb{A}^{\rm d}\mb{\lambda}^{\rm d}\tilde{\!\bar{F}} \bar{H}\; .
\end{equation}
Universality at the Planck scale is assumed:
\begin{equation}\label{eq:m0A0}
\begin{array}{c}
\mb{m}_T^2 = \mb{m}_{\bar{F}}^2 = m_0^2 \uno,\\[1mm]
m_H^2=m_{\bar H}^2 = m_0^2,\\[1mm]
\mb{A}^{\rm u} = \mb{A}^{\rm d} = A_0 \uno.
\end{array}\end{equation}
By solving the RGE's, 
one finds at $M_{G}$ \cite{barbieri}:
\begin{eqnarray}
\mb{m}_{T\rm G}^2 &=&
\diag(m_{T\rm G}^2,m_{T\rm G}^2,m_{T\rm G}^2-I_{\rm G})
\equiv m_{T\rm G}^2\uno - \I_{\rm G},\nn\\
\mb{A}^{\rm d}_{\rm G} &=&\textstyle
\diag(A_{d\rm G},A_{d\rm G},A_{d\rm G}- \frac{1}{3}I'_{\rm G})
\equiv A_{d\rm G}\uno-\frac{1}{3}\I'_{\rm G},\nn\\
\mb{A}^{\rm u}_{\rm G} &=&\textstyle
\diag(A_{u\rm G}-\frac{1}{3}I'_{\rm G},A_{u\rm G}-\frac{1}{3}I'_{\rm G}
,A_{u\rm G}-I'_{\rm G}),\nn\\
\mb{m}_{\bar{F}\rm G}^2 &=& \textstyle
m_{\bar{F}\rm G}^2\uno,\nn\\
\mb{\lambda}^{\rm u}_{\rm G}&=& \textstyle
\diag(\lambda_{u\rm G},\lambda_{c\rm G},\lambda_{t\rm G}),
\label{mgsu5}
\end{eqnarray}
whereas $\mb{\lambda}^{\rm d}$ gets renormalized
to      $\mb{\lambda}^{\rm d}_{\rm G}$. The explicit expression of all the 
quantities appearing in eq.~(\ref{mgsu5}) can be found in ref.~\cite{barbieri}.
From this computation one sees that the large value of the top Yukawa coupling, 
$\lambda_{t}$, produces major effects on the non-universality of the soft
breaking terms after renormalizing them from $M_{\rm Pl}$ to $M_{\rm G}$.

After SU(5) breaking one obtains the following slepton and lepton mass 
matrices at the
scale $M_{Z}$:
\begin{equation}\label{eq:lowEnLag}
-\cal{L}_m^{\rm sl}=
\tilde{L}^\dagger \mb{m}_L^2\tilde{L} +
\tilde{e}_L^{c\dagger} \mb{m}_e^2 \tilde{e}_L^{c} +
\tilde{e}_L^{cT} (\mb{A}^{\rm e} +  \uno\mu\tan\beta)\mb{\lambda}^{\rm e}
\tilde{e}_L v_{\rm d}+{\rm h.c.}
\end{equation}
where 
\begin{equation}
\mb{m}_L^2=m_L^2\uno,\qquad
\mb{m}_e^2=m_e^2\uno-\I_{\rm G},\qquad
\mb{A}^{\rm e}=A_e\uno-{\textstyle\frac{1}{3}}\I'_{\rm G},
\end{equation}
in a self-explanatory notation,
\begin{equation}
\cal{L}_Y=
     Q^T\mb{\lambda}^{\rm u}_Z u^c_L\cdot v_{\rm u} +
     Q^T\mb{\lambda}^{\rm d}_Z d^c_L\cdot v_{\rm d} +
e^{cT}_L\mb{\lambda}^{\rm e}_Z L    \cdot v_{\rm d}
\end{equation}
where $\mb{\lambda}^{\rm u}_Z$ has kept
its diagonal form and the matrices
$\mb{\lambda}^{\rm d}$ and $\mb{\lambda}^{\rm e}$,
equal at $M_{\rm G}$, have been shifted by the different renormalization
effects due to $\lambda_t$ and the gauge couplings.
By diagonalizing $\mb{\lambda}^{\rm d}_Z$ and
$\mb{\lambda}^{\rm e}_Z$, we have
\begin{eqnarray}
\mb{\lambda}^{\rm d}_Zv_{\rm d}&=&\mb{V}^* \mb{M}^{\rm d}\mb{U}^\dagger\nn\\
\mb{\lambda}^{\rm e}_Zv_{\rm d}&=&\mb{V}^{\rm e*}\mb{M}^{\rm e}
\mb{U}^{\rm e\dagger}\label{eq:Ue}
\end{eqnarray}
where $\mb{M}^{\rm d}$, $\mb{M}^{\rm e}$ are the diagonal mass
matrices for down quarks and charged leptons,
$\mb{U}=\mb{U}^{\rm e}$,
$\mb{V}$ is the usual Cabibbo-Kobayashi-Maskawa matrix and,
as an effect of the top Yukawa coupling, the matrix elements of
$\mb{V}^{\rm e}$ are related to those of $\mb{V}$ by
\begin{equation}\label{eq:Vscaling}
V^{\rm e}_{ij}=y V_{ij}\quad\hbox{for $i\neq j$ and ($i$ or $j)=3$},
\qquad
V^{\rm e}_{ij}= V_{ij}\quad{\mbox{otherwise}}
\end{equation}
and $y$ is defined in ref.~\cite{barbieri}.

We now switch to a mass eigenstate basis for the charged leptons:
\begin{equation}\label{eq:ridefl}
\mb{V}^{\rm e\dagger} e^c_L = e^{c \prime}_L,\qquad
\mb{U}^{\rm e\dagger} L = L'.
\end{equation}
In order to keep neutral vertices diagonal in flavour space, we rotate 
sleptons simultaneously with leptons (we suppress primes after eq. 
(\ref{eq:ridefsl})):
\begin{equation}\label{eq:ridefsl}
\mb{V}^{\rm e\dagger} \tilde{e}^c_L = \tilde{e}^{c \prime}_L,\qquad
\mb{U}^{\rm e\dagger} \tilde{L} = \tilde{L}'.
\end{equation}
This generates off-diagonal slepton mass terms:
\begin{eqnarray}
\left(\delta_{ij}^{l}\right)_{RR}&=&-\mb{V}^{\rm e *}_{3i}\mb{V}^{\rm e}_{3j}
\frac{I_{\rm G}}{m^{2}_{\tilde{l}}} \nn \\
\left(\delta_{ij}^{l}\right)_{RL}&=&-\frac{1}{3}\mb{V}^{\rm e }_{3i}
\mb{V}^{\rm e*}_{3j} 
\frac{M^{\rm e}_{j}I^{\prime}_{\rm G}}{m_{\tilde{l}}^{2}}\; ,
\label{deltaijsu(5)}
\end{eqnarray}
where we have defined as usual an average slepton mass $m_{\tilde{l}}$.

We now note that for large $\tan\beta$ it may be convenient to perform a 
Left-Right transition via a double mass insertion, a Left-Right flavour 
diagonal one $\left(\delta_{ii}\right)_{LR}=M^{e}_{i}(A_{e}+\mu \tan \beta)/
m_{\tilde{l}}^{2}$
 followed by  a 
Right-Right flavour 
changing one $\left(\delta_{ij}\right)_{RR}$. Including this possibility 
one gets the following general expression for the decay $l_i \to \l_j \gamma$:
(see ref.~\cite{gabbmas}):
\begin{eqnarray}
	{\mbox{BR}}(l_{i} \to l_{j} \gamma) &=& \frac{\alpha^{3}}{G_{F}^{2}}
	\frac{12 \pi} 
	{m_{\tilde{l}}^{4}}\left\{\left\vert \left[M_{3}(x)+\frac{3}{2}
	\left(A_{\rm e} + \mu \tan\beta \right) 
	\frac{m_{\tilde{\gamma}}}{m_{\tilde{l}}^{2}} \tilde{f}_{6}(x) \right]
	\left(\delta^{l}_{ij}\right)_{LL} \right. \right. 
	\nonumber \\
	&&\left. \left. + \frac{m_{\tilde{\gamma}}}{m_l} 
	M_{1}(x) \left(\delta^{l}_{ij}\right)_{LR}\right\vert^{2}
	+ L \leftrightarrow R \right\} \cdot 
	{\mbox{BR}}(l_{i} \to l_{j} \nu_{i} \bar{\nu}_{j}) \; .
	\label{litolj2}
\end{eqnarray}
In tables \ref{tab:su(5)1} and \ref{tab:su(5)2}, 
we compare the limits, obtained 
imposing that each individual term in eq.~(\ref{litolj2}) does not exceed the 
value given in table~\ref{ul}, with the values predicted in SU(5).
\begin{table}
	\begin{center}
	\begin{tabular}{||c|c|c|c|c||}
	\hline \hline
	& & & &\\
	$y$ & $\left| \left(\delta^l_{12} \right)_{RR} \right|^{ex}$ 
	& $\left| \left( \delta^l_{12} \right)_{RR} \right|^{th}$ 
	& $\left| \left( \delta^l_{12} \right)_{RL} \right|^{ex}$ 
	& $\left| \left( \delta^l_{12} \right)_{RL} \right|^{th}$ \\
	& & & & \\
	\hline
	0.3 & $4.3 \times 10^{-3}$ & $1.8 \times 10^{-4}$ 
	& $1.5 \times 10^{-6}$ & $1.0 \times 10^{-7}$ \\
	0.7 & $6.5 \times 10^{-3}$ & $1.5 \times 10^{-4}$ 
	& $1.6 \times 10^{-6}$ & $1.1 \times 10^{-7}$ \\
	1.0 & $8.3 \times 10^{-3}$ & $1.2 \times 10^{-4}$ 
	& $1.8 \times 10^{-6}$ & $1.2 \times 10^{-7}$ \\
	\hline \hline
	\end{tabular}
	\caption[]{Comparison between experimental limits 
	$\vert \left( \delta^l_{ij} \right)_{AB} \vert^{ex}$ and values 
	obtained in SU(5) for the $\delta$'s. We have used 
	$m_{\tilde{e}_{R}}=100$ 
	GeV, $\lambda_{tG}=1.4$, $A_e/m_{\tilde{e}_{R}}=-1$, $\mu < 0$
	and different values of $y=m_{\tilde{\gamma}}^2/m_{\tilde{e}_{R}}^2$.}
	\protect\label{tab:su(5)1}
 	\end{center}
\end{table}
\begin{table}
	\begin{center}
		\begin{tabular}{||c|c|c|c||}
		\hline \hline
		& & & \\ 
		$y$ & 
		$\left| \left( \delta^l_{12} \right)_{RR}
		\left( \delta^l_{22} \right)_{RL} \right|^{ex}$ 
		& $\left| \left( \delta^l_{12} \right)_{RR}
		\left( \delta^l_{22} \right)_{RL} 
		\right|^{th}$  
		& $\left| \left( \delta^l_{12} \right)_{RR}
		\left( \delta^l_{22} \right)_{RL} 
		\right|^{th}$ \\ 
		& & {\small ($\tan \beta=2$)} & {\small ($\tan \beta=10$)}\\
		\hline
		0.3 & $2.6 \times 10^{-6}$ 
		& $7.1 \times 10^{-7}$ & $2.2 \times 10^{-6}$ \\ 
		0.7 & $3.3 \times 10^{-6}$ 
		& $7.5 \times 10^{-7}$ & $2.7 \times 10^{-6}$  \\
		1.0 & $4.5 \times 10^{-6}$ 
		& $7.0 \times 10^{-7}$ & $2.6 \times 10^{-6}$  \\
		\hline \hline
		\end{tabular}
	\caption[]{Same as in table~\protect\ref{tab:su(5)1} for double mass 
	insertions.}
	\protect\label{tab:su(5)2}
	\end{center}
\end{table}	
\subsection{SUSY SO(10)}
Following the same procedure as in Section~\ref{sec:su(5)}, one can derive the 
expression for the off-diagonal slepton mass terms. We maintain here the same 
notation as in Section~\ref{sec:su(5)}, bearing in mind that the relations 
between low- and high-energy parameters have now changed (we refer again the 
reader to ref.~\cite{barbieri} for details).
One gets
\begin{eqnarray}
\left(\delta_{ij}^{l}\right)_{LL}&=&-\mb{V}^{\rm e}_{3i}\mb{V}^{\rm e*}_{3j}
\frac{I_{\rm G}}{m^{2}_{\tilde{l}}} \nn \\
\left(\delta_{ij}^{l}\right)_{RR}&=&-\mb{V}^{\rm e *}_{3i}\mb{V}^{\rm e}_{3j}
\frac{I_{\rm G}}{m^{2}_{\tilde{l}}} \nn \\
\left(\delta_{ij}^{l}\right)_{LR}&=&-\frac{5}{7}\left[\mb{V}^{\rm e }_{3i}
\mb{V}^{\rm e*}_{3j}M^{\rm e}_{j} + \mb{V}^{\rm e *}_{3i}\mb{V}^{\rm e}_{3j} 
M^{\rm e}_{i}\right]\frac{I^{\prime}_{\rm G}}{m_{\tilde{l}}^{2}}\; .
\label{deltaijso(10)}
\end{eqnarray}
We now note that flavour violating terms which mediate the transition from 
flavour $i$ to flavour $j$ are always proportional to $\mb{V}^{\rm e}_{3i}$ 
and to $\mb{V}^{\rm e}_{3j}$. Therefore, a $1-2$ transition must be 
proportional  to $\mb{V}^{\rm e*}_{31} \mb{V}^{\rm e}_{32}$. On the other 
hand, an $i-3$ transition is only proportional to $\mb{V}^{\rm e}_{3i}$, as 
$\mb{V}^{\rm e}_{33}=1$. This means that a double mass insertion in which the 
intermediate flavour index is $3$ is not suppressed with respect to a single 
insertion, as long as $I_{\rm G}/m^{2}_{\tilde{l}}$ is of order one. 
In this case, by performing a double mass insertion, one can obtain an 
amplitude for the decay $\mu \to e \gamma$ which is proportional to $m_{\tau}$ 
instead of $m_{\mu}$.  If one takes this into account, the mass insertion 
method gives a good approximation of the complete result, whereas if one 
ignores the possibility of such double mass insertions, the method gives a 
poor approximation, as noted in ref.~\cite{barbieri}. 

As in SU(5), in the large $\tan\beta$ region it may be 
convenient to perform a Left-Right transition via 
the flavour conserving $M^{\rm e}\left(A_{\rm e} + \mu \tan\beta \right)$ 
term, followed by an $LL$ or $RR$ flavour changing mass insertion. Hence, 
given that this latter term can be itself obtained by a double mass insertion,
we end up with a competitive triple mass insertion.
We have 
approximated the triple mass insertion by a double one with an {\em effective} 
Left-Right mass insertion given by 
\begin{equation}
\left(\delta_{i3}^{l}\right)^{effective}_{LR}=\left(\delta_{i3}^{l}\right)_{LL}
\times 
\frac{M^{\rm e}_{3}\left(A_{\rm e} + \mu \tan\beta \right)}{m_{\tilde{l}}^{2}}
\; .
\label{effective}
\end{equation}
A comparison between experimental limits and theoretical predictions is given 
in tables \ref{tab:so(10)1} and \ref{tab:so(10)2}.  
\begin{table}
	\begin{center}
	\begin{tabular}{||c|c|c|c|c||}
		\hline \hline
		& & & & \\
		$y$ & 
		$\left| \left(\delta^l_{12} \right)_{RR} \right|^{ex}$ 
		& $\left| \left( \delta^l_{12} \right)_{RR} \right|^{th}$ 
		& $\left| \left( \delta^l_{12} \right)_{RL} \right|^{ex}$ 
		& $\left| \left( \delta^l_{12} \right)_{RL} \right|^{th}$ \\
		& & & & \\
		\hline
		0.1 & $2.8 \times 10^{-2}$ & $2.1 \times 10^{-4}$ 
		& $4.7 \times 10^{-6}$ & $3.6 \times 10^{-8}$ \\
		0.3 & $3.9 \times 10^{-2}$ & $1.9 \times 10^{-4}$ 
		& $4.4 \times 10^{-6}$ & $4.2 \times 10^{-8}$ \\
		0.7 & $6.0 \times 10^{-2}$ & $1.4 \times 10^{-4}$ 
		& $5.0 \times 10^{-6}$ & $4.7 \times 10^{-8}$ \\
		\hline \hline
	\end{tabular}
	\caption[]{Comparison between experimental limits 
	$\vert \left( \delta^l_{12} \right)_{AB} \vert^{ex}$ and values 
	obtained in $SO(10)$ for the $\delta$'s. 
	In this case the model is $L-R$ 
	symmetric, and limits are independent from the exchange $L 
	\leftrightarrow R$. We have used $m_{\tilde{e}_{R}}=300$ 
	GeV, $\lambda_{tG}=1.25$, $A_e/m_{\tilde{e}_{R}}=-1$, $\mu <0$
	and different values of $y=m_{\tilde{\gamma}}^2/m_{\tilde{e}_{R}}^2$.}
	\protect\label{tab:so(10)1}
 	\end{center}
\end{table}
\begin{table}
	\begin{center}
		\begin{tabular}{||c|c|c|c|c||}
		\hline \hline
		& & & & \\
		$y$ & 
		$\left| \left( \delta^l_{13} \right)_{RR}
		\left( \delta^l_{32} \right)_{RL} \right|^{ex}$ 
		&  $\left| \left( \delta^l_{13} \right)_{RR}
		\left( \delta^l_{32} \right)_{RL} 
		\right|^{th}$  
		&  $\left| \left( \delta^l_{13} \right)_{RR}
		\left( \delta^l_{32} \right)_{RL}^{eff} 
		\right|^{th}$  
		&  $\left| \left( \delta^l_{13} \right)_{RR}
		\left( \delta^l_{32} \right)_{RL}^{eff} 
		\right|^{th}$ \\ 
		& & &  ($\tan \beta=2$) &  ($\tan \beta=10$)\\
		\hline
		0.1 & $4.5 \times 10^{-6}$ & $1.8 \times 10^{-5}$ 
		& $7.5 \times 10^{-5}$ & $1.9 \times 10^{-4}$  \\ 
		0.3 & $6.4 \times 10^{-6}$ & $1.3 \times 10^{-5}$ 
		& $4.7 \times 10^{-5}$ & $1.5 \times 10^{-4}$  \\
		0.7 & $9.6 \times 10^{-6}$ & $7.2 \times 10^{-6}$ 
		& $1.8 \times 10^{-5}$ & $6.4 \times 10^{-5}$  \\
		\hline \hline
		\end{tabular}
	\caption[]{Same as in table~\protect\ref{tab:so(10)1} for double mass 
	insertions. In this case the relevant average slepton mass is defined 
	as $m_{\tilde{l}}=\sqrt{m_{\tilde{e}_R}m_{\tilde{\tau}_R}}$. 
	The last two columns contain effective insertions, as 
	defined in the text.}
	\protect\label{tab:so(10)2}
	\end{center}
\end{table}	
\subsection{Non-universal soft breaking terms}
\begin{figure}   
    \begin{center}
    \epsfysize=12truecm
    \leavevmode\epsffile{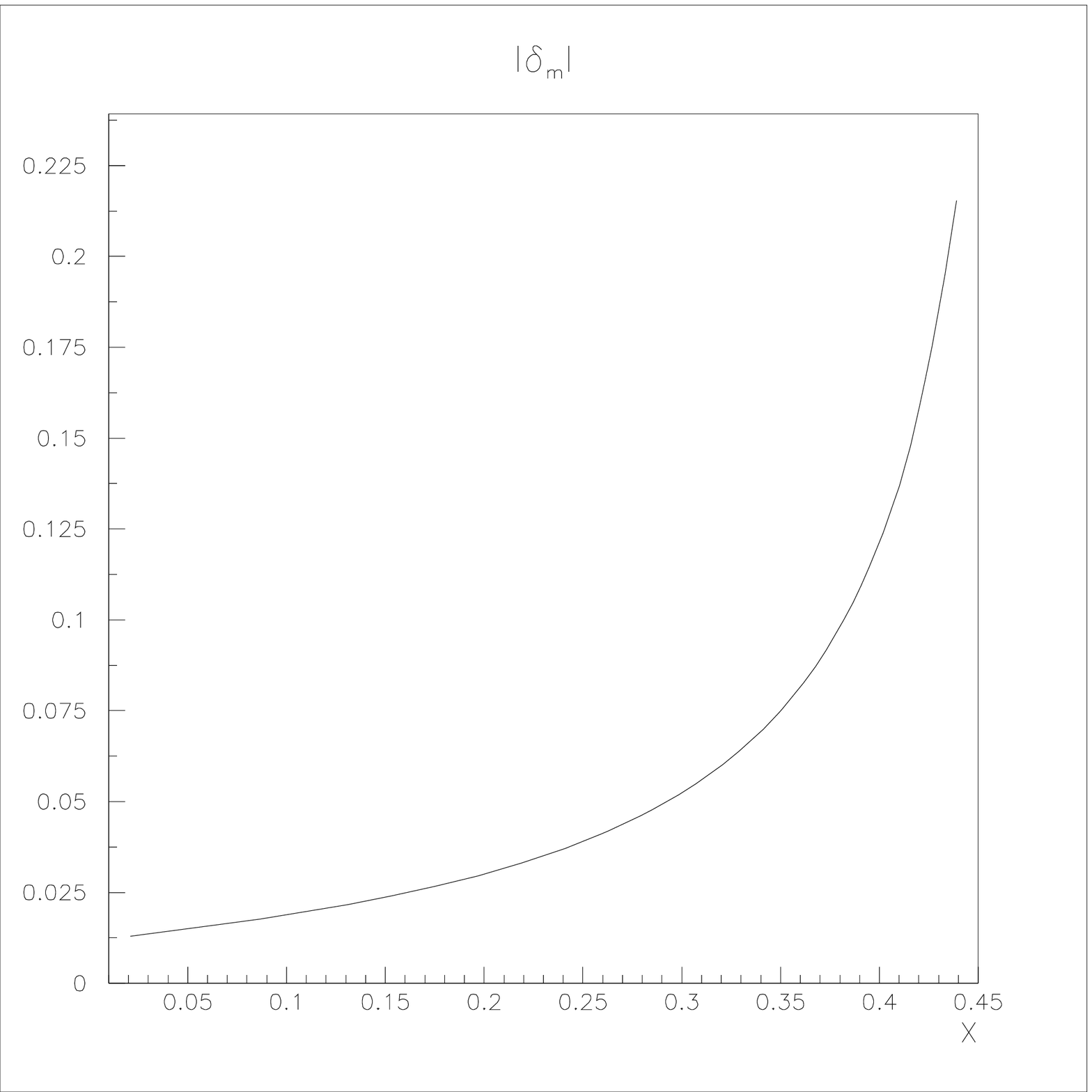}
    \end{center}
    \caption[]{The $\vert\delta_{m}\vert$ as a function
     of $x=m_{\tilde{\gamma}}^2/m_{\tilde{l}}^2$, for  an average slepton mass 
     $m_{\tilde{l}}=100\mbox{GeV}$.}
    \protect\label{fig:delta01l}
\end{figure}

We now consider a simple model with minimal non-universality in the leptonic 
sector. Let us assume that the soft breaking terms for sleptons at the GUT 
scale $M_{G}$ have the following form:
\begin{equation}\label{eq:Vsoftl}
V_{\rm soft}=
      \tilde{l}_{L}^\dagger \mb{m}_l^2\tilde{l}_{L}+
\tilde{e}_{L}^{c\dagger} \mb{m}_{e}^2 \tilde{e}_{L}^{c}+
\tilde{e}_{L}^{cT} \mb{A}^{\rm l}\mb{\lambda}^{\rm l}\tilde{l}_{L} \bar{H}\; ,
\end{equation}
where 
\begin{eqnarray}
\mb{m}^{2}_l&=&\diag(\tilde{m}_0^{2}+\Delta_{m^2},\tilde{m}^{2}_0,
\tilde{m}^{2}_0-\Delta_{m^2})\nn  \\
\mb{m}^{2}_{e}&=&\tilde{m}_0^{2} \uno \nn \\
\mb{A}^{\rm l}&=&A_{l}\uno.
\end{eqnarray}
We now assume for simplicity that the Yukawa couplings of leptons are 
proportional to the ones of $d$-quarks in the basis where the couplings of 
$u$-quarks are diagonal. Performing the RGE evolution down to the electroweak 
scale, diagonalizing the lepton mass matrix and rotating sleptons to keep the 
$l - \tilde{l}-\tilde{\gamma}$ vertex diagonal, we get the flavour-violating 
mass insertion between selectrons and smuons
\begin{equation}
\left(\delta^{l}_{12}\right)_{LL}=\left(\mb{V}^{\rm l}_{11}\mb{V}^{\rm l*}_{12}
-\mb{V}^{\rm l}_{13}\mb{V}^{\rm l*}_{32}\right)\frac{\Delta_{m^2}}
{\tilde{m}^{2}}
\simeq \mb{V}^{\rm }_{11}\mb{V}^{\rm *}_{12}\frac{\Delta_{m^2}}
{\tilde{m}^{2}}\; ,
\label{deltal}
\end{equation}
where $\mb{V}$ is the CKM matrix and $\tilde{m}$ is the low-energy slepton 
average mass.
Starting from the limits in table \ref{lep} we obtain the constraints on 
$\delta_m=\Delta_{m^2}/\tilde{m}^{2}_0$ 
plotted in figure \ref{fig:delta01l}, as a function of $x=
m^{2}_{\tilde{\gamma}}/m^{2}_{\tilde{l}}$. Notice that, differently from our 
previous discussion, here we restrict to the range $x < 0.5$. This is due to 
the fact that, as shown in ref.~\cite{pokorski}, the running of the mass 
parameters implies an upper bound on the ratio $m^{2}_{\tilde{\gamma}}/
m^{2}_{\tilde{l}}$ at the electroweak scale of order 0.5. 

If one compares the results plotted in fig.~\ref{fig:delta01l} with those in 
table~\ref{lep}, one finds that the ``dilution'' of the degeneracy constraint 
when going from the low to the large scale increases for a more accentuated 
gaugino dominance. Namely, the larger $x$ is (compatibly with $x<0.5$), the 
weaker is the constraint on $\delta_{m}$. 
\section{Conclusions and outlook}
\label{concl}
In this paper we have provided a systematic study of all the most stringent 
constraints coming from FCNC and CP violating phenomena on the off-diagonal 
flavour-changing terms in the sfermion mass matrices. Our model-independent 
parameterization, which makes use of the mass-insertion method, is 
particularly suitable for a ready check of the viability of any SUSY extension 
of the SM in view of the FCNC and CP tests. Obviously, such a kind of check 
can be considered only as a first, coarse approach to the full analysis of the 
FCNC and CP predictions within a specific SUSY model which requires the full 
diagonalization of the sfermion mass matrices.
Needless to say, we think that the basic ignorance of the mechanism 
responsible for SUSY breaking and the consequent large variety of SUSY models 
in the context of supergravity and superstring theories 
make it necessary to have an adequate, preliminary test of the 
crucial FCNC and CP quantities readily available. Our analysis clarifies the 
extent to which the study of FCNC and CP via the mass insertion method is 
valid. In particular, the Section devoted to SUSY-GUT's makes it clear that a 
naive implementation of the method may lead to results which substantially 
differ from what is obtained in the ``correct'' mass eigenstate formalism.

A legitimate question that one can formulate at the end of such a long 
analysis is whether we can hope to find some indirect manifestation of SUSY 
through FCNC and CP violating phenomena and, if so, what  are the best 
candidates. Obviously the answer is highly model-dependent. For instance, the 
MSSM might easily escape any kind of indirect manifestation through FCNC and/
or CP violation. On the other hand, SUSY-GUT's and models with 
non-universality have a high potentiality to produce FCNC and/or CP violating 
phenomena at a rate which is experimentally detectable. If a unification of 
leptons and quarks into common multiplets occurs, and if one can trust the use 
of RGE's in a delicate range such as that between $M_{\rm Pl}$ and $M_{\rm 
G}$, then phenomena as $\mu \to e \gamma$ or $\mu-e$ conversion in atoms are 
the most likely candidates to exhibit some signal of new physics. In the 
sector of CP violation, as we have seen, if the $\delta_{LR}$ insertions are 
proportional to the Yukawa couplings of the d- and s-quarks, then $\delta_{LR} 
\ll \delta_{LL}$ and the SUSY contribution from gluino exchange is essentially 
of superweak nature. On the contrary, if one envisages the presence of a 
conspicuous $\delta_{LR}$ in the kaon system, then, while respecting the bound 
from $\varepsilon$, it is possible to obtain large SUSY contributions to 
$\varepsilon^\prime/\varepsilon$. 
Notice, however, that $\Im \left(\delta^{d}_{11}\right)_{LR}$ is strongly 
constrained by $d_{n}$ (see table~\ref{dipoles}). Unless $\Im \left(
\delta^{d}_{12}\right)_{LR} \gg \Im \left(
\delta^{d}_{11}\right)_{LR}$ there is no hope for a sizeable SUSY contribution 
to $\varepsilon^{\prime}/\varepsilon$ even in models where the $\delta_{LR}$ 
quantities are not proportional to Yukawa couplings.
Finally, B physics can be quite sensitive to 
the presence of gluino-mediated SUSY contributions. While the B oscillations 
can receive sizeable contributions both with or without a conspicuous 
$\delta_{LR}$, this mass insertion is crucial if one asks for enhancement in 
the process $b \to s \gamma$ which requires a helicity flip. There is also a 
potentiality of conspicuous contributions to CP violation in B physics which 
has not been explored here.

In conclusion, our analysis confirms a twofold role played by FCNC and CP 
violating phenomena in shedding some light on the SUSY extensions of the SM. On 
one side, our study emphasizes that the constraints coming from this class of 
processes are very severe and impose rather stringent selections of 
fundamental theories whose low energy limit is a SUSY extension of the SM. On 
the other hand, it emerges from
our analysis that the bounds on the $\delta$ quantities that one derives from 
the available experimental data are not far from (or, even, clash with) the 
values for the $\delta$'s that one finds in effective N=1 supergravities or 
SUSY-GUT's.
In view of this latter observation, it 
is not unconceivable that, after all, SUSY may manifest itself through its 
contributions to FCNC and/or CP violating processes even before its direct 
discovery through the production of SUSY particles.
\section*{Acknowledgements}
We thank R. Barbieri and R. Petronzio for useful discussions. Two of us (E.G. 
and L.S.) are grateful to the physics departments of Padova and Perugia for 
their kind hospitality during the completion of this work.
E.G. would like to thank the theory division of CERN and the physics 
department of Southampton for their warm hospitality.
\renewcommand{\baselinestretch}{1}

\newpage
\newpage
\newpage
\newpage
\newpage
\begin{figure}   
    \begin{center}
    \epsfysize=8truecm
    \leavevmode\epsffile{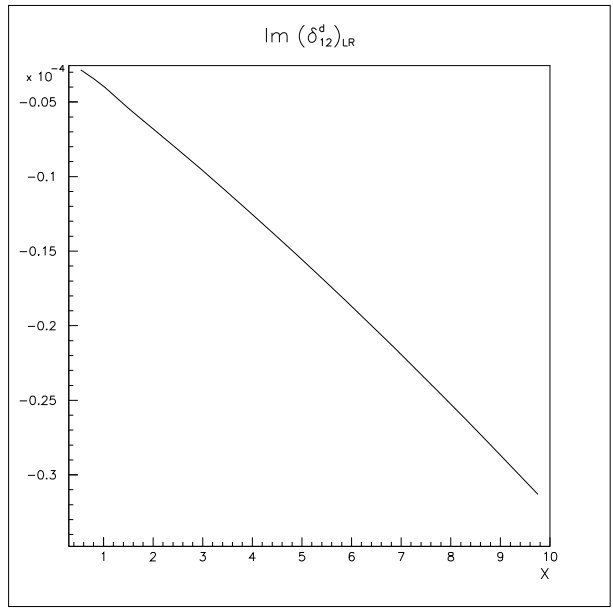}
    \end{center}
    \caption[]{The $\mbox{Im}\left(\delta^d_{12}\right)_{LR}$ as a function
     of $x=m_{\tilde{g}}^2/m_{\tilde{q}}^2$, for  an average squark mass 
     $m_{\tilde{q}}=100\mbox{GeV}$.}
    \label{epplr}
\end{figure}
\end{document}